\documentclass[12pt, draftclsnofoot, onecolumn]{IEEEtran}
\pdfoutput=1
\usepackage{color}
\usepackage{soul} 
\usepackage{cancel} 
\ifCLASSINFOpdf
\else
\fi
\usepackage{graphicx}
\graphicspath{{figures/A_complexity/}{figures/B_exemple/}{figures/C_markov_model/}{figures/D_model_comparison_mc/A_TEL/}{figures/D_model_comparison_mc/B_ODL/}{figures/E_comparison_with_Rose/}{figures/F_comparaison_theorique/}}
\DeclareGraphicsExtensions{.pdf}

\newtheorem{definition}{Definition}

\newtheorem{hypothesis}{Hypothesis}

\newtheorem{observation}{Observation}

\newtheorem{ass}{Assumption}

\usepackage{amsmath}
\usepackage{amsfonts} 
\usepackage{dsfont}
\usepackage[ruled]{algorithm}
\usepackage{algpseudocode}

\newcommand{\E}[2]{\mathds{E}_{#2}\left[ #1 \right]}
\newcommand{\ind}[1]{\mathds{1}_{\left\lbrace #1 \right\rbrace}}

\newcommand{\Prob}[1]{\text{Pr}\left\{ #1 \right\}}
\newcommand{\ac}[1]{\ac{ #1} }
\newcommand{\bz}{\mathbf{z}}
\newcommand{\bP}{\mathbf{P}}
\newcommand{\bS}{\mathbf{S}}
\newcommand{\bZ}{\mathbf{Z}}
\newcommand{\bF}{\mathbf{F}}
\newcommand{\bI}{\mathbf{I}}
\newcommand{\bOne}{\mathbf{1}}
\newcommand{\bb}{\mathbf{b}}
\newcommand{\bm}{\mathbf{m}}
\newcommand{\ba}{\mathbf{a}}
\newcommand{\bu}{\mathbf{u}}
\newcommand{\bs}{\mathbf{s}}
\newcommand{\bd}{\mathbf{d}}
\newcommand{\bpi}{\mathbf{\pi}}
\newcommand{\bba}{\mathbf{\bar{a}}}
\newcommand{\bbu}{\mathbf{\bar{u}}}
\newcommand{\Players}{\mathcal{K}}
\newcommand{\Channels}{\mathcal{N}}
\newcommand{\Game}{\mathcal{G}}
\newcommand{\Actionset}{\mathcal{A}}
\newcommand{\cR}{\mathcal{R}}

\newcommand{\cS}{\mathcal{S}}
 
\usepackage[shortcuts,acronym,toc]{glossaries} 
\makeglossaries
\newacronym{TE}{LLTEL}{Log Linear Trial and Error Learning}
\newacronym{MTE}{MTEL}{Modified Trial and Error Learning}
\newacronym{TEL}{TEL}{Trial and Error Learning}
\newacronym{ODL}{ODL}{Optimal Distributed Learning}
\newacronym{MODL}{MODL}{Modified Optimal Distributed Learning}
\newacronym{FSC}{FSC}{Finite State Controller}
\newacronym{NE}{NE}{Nash Equilibrium}
\newacronym{PNE}{PNE}{Pure Nash Equilibrium}
\newacronym{CH}{CH}{player Head}
\newacronym{SINR}{SINR}{Signal to Noise plus Interference Ratio}
\newacronym{NB}{NB}{Narrow Band}
\newacronym{WB}{WB}{Wideband}
\newacronym{SSS}{SSS}{Stochastic Stable State}
\newacronym{RL}{RL}{Reinforcement Learning}
\newacronym{RRC}{RRC}{Reduced Recurrent Classes}
\newacronym{MC}{MC}{Markov Chain}
\newacronym{EFHT}{EFHT}{Expected First Hitting Time}
\newacronym{AODL}{AODL}{Adapted ODL}
\newacronym{RC}{RC}{{\it recurrence classes}}

\ifCLASSOPTIONcompsoc
  \usepackage[caption=false,font=normalsize,labelfont=sf,textfont=sf]{subfig}
\else
  \usepackage[caption=false,font=footnotesize]{subfig}
\fi
\begin{document}
%
\title{Performance Analysis of Trial and Error Algorithms}
%
%
%

\author{J\'er\^ome~Gaveau, \IEEEmembership{Student Member,~IEEE,}
        Christophe~J.~Le~Martret,~\IEEEmembership{Senior Member,~IEEE}
        and~Mohamad~Assaad,~\IEEEmembership{Senior Member,~IEEE,}
        }

\maketitle
\begin{abstract}
	
Model-free decentralized optimizations and learning  are receiving increasing attention from theoretical and practical perspectives. In particular, two fully decentralized learning algorithms, namely Trial and Error (TEL) and Optimal Dynamical Learning (ODL), are very appealing for a broad class of games. In fact, ODL has the property to spend a high proportion of time in an optimum state that maximizes the sum of utility of all players. And the TEL has the property to spend a high proportion of time in an optimum state that maximizes the sum of utility of all players if there is a Pure Nash Equilibrium (PNE), otherwise, it spends a high proportion of time in an optimum state that maximizes a tradeoff between the sum of utility of all players and a predefined stability function. On the other hand, estimating the mean fraction of time spent in the optimum state (as well as the mean time duration to reach it) is challenging due to the high complexity and dimension of the inherent Markov Chains. In this paper, under some specific system model,  an evaluation of the above performance metrics is provided by proposing an approximation of the considered Markov chains, which allows overcoming the problem of high dimensionality.  A comparison between the two algorithms is then performed which allows a better understanding of their performances. 
\vspace{-0.5cm}
\end{abstract}


%
\IEEEpeerreviewmaketitle

\section{Introduction}
\label{sec:introduction}
Game Theory and more generally decentralized optimization has recently received increasing attention from theoretical and practical perspectives. For instance, several classes of games have been studied and characterization of the corresponding  equilibria has been performed \cite{fudenberg1998theory,lasaulce2011game}.  On another hand, developing learning based methods that can be implemented in a distributed way by the agents is of paramount importance in decentralized optimization and game frameworks. These methods must either converge to an equilibrium, in game contexts,  or to a local/global optimum in the context of decentralized optimization. One can refer to \cite{fudenberg1998theory} for a survey on learning based methods.

In most cases, it is assumed that the utility function of the users and action set have some mathematical properties (e.g. Lipschitz continuity of the reward, etc.) to ensure the convergence of  developed methods.

In practice, the optimization/game frameworks can however very complex, in which the utility function may not have a closed form expression and even may take discrete values. In such contexts, model-free strategy learning algorithms are very appealing approaches \cite{wang2016surveymodelfree}. Players neither try to model the environment nor try to have a specific/explicit  utility form. They simply consider the environment  as a black box and learn by interactions (e.g. trials and errors). This context, though very restrictive, can be encountered in a wide variety of examples. For instance, in a wind farm, each turbine controls the power that it extracts from the wind \cite{marden2014achieving}. It is very difficult, if not intractable, to model the impact of a turbine on other turbines. In addition, the lack of communications  between them makes impossible any cooperation. Another example is the case of commuters in city that want to avoid traffic jams but, they neither know the strategies of other commuters  nor the impact of their strategy on the achieved rewards \cite{young1998learningTE}. In the context of wireless  telecommunication systems, decentralized resource allocation can be encountered in many contexts since the nodes/players may not be able to exchange information between each other in order not to increase the overhead in the network. Also realistic  utility functions of the users may not have closed form expression (e.g. Quality of Experience, number of correctly decoded packets; etc.). Decentralized resource allocation approaches have been used \cite{bennis2013self,perlaza2010justerl} to respectively share the resources  among femtocells or wifi access points.   In ad hoc networks, the network is infrastructure-less which makes decentralized learning solutions suited in such contexts \cite{rose2012startTE,rose2014self,sheng2014utility,rose2013achieving}.

In model-free resource allocation schemes, developing decentralized strategies that converge to an equilibrium (if it exists), or at least finding conditions under which they converge, represent a main challenge \cite{wang2016surveymodelfree}. The trial and error algorithms, proposed in \cite{marden2014achieving,pradelski2012learning} and then applied to various resource sharing problems e.g. \cite{rose2012startTE,rose2014self,sheng2014utility,rose2013achieving}, are very appealing in these contexts. They show the particularity to exhibit cooperative convergence properties in a broad class of games. For this reason, the focus in this paper is on Trial and Errors algorithms. For instance, \ac{ODL} algorithm from \cite{marden2014achieving} has the property to spend a high proportion of time in an optimum state that maximizes the sum of utility of all players whether there is, or not, a \ac{PNE}. On the other hand, \ac{TEL} algorithm from \cite{pradelski2012learning} has the property to spend a high proportion of time in an optimum state that maximizes the sum of utility of all players if there is a \ac{PNE}, otherwise, it spends a high proportion of time in an optimum state that maximizes a tradeoff between the sum of utility of all players and a predefined stability function. 
Even though the above two algorithms converge to a desired state, the convergence rate remains an open question \cite{wang2016surveymodelfree,pradelski2012learning}. The main reason comes from the computation complexity of the inherent \ac{MC} generated by these two algorithms. In fact, the game in which players employ these learning schemes can be represented by discrete \ac{MC}s with huge number of states. Obtaining the transitions matrix of  these \ac{MC}s is therefore not tractable which makes the analysis of the convergence rate not possible (even numerically). 

The main contributions of this work are fourfold. We are interested in computing the mean time these algorithms spend in a desired state as well as the mean time required to achieve that state under a given model. Due to the huge dimension of the \ac{MC}s, only approximations can be employed to compute a close approximation of the aforementioned convergence metrics. The first contribution is to provide an approximation of the \ac{MC} associated to the \ac{TEL} algorithm. The second contribution is to also provide such an approximation for \ac{ODL} algorithm. In addition, we explain the methodology to obtain them. To the best of our knowledge, a first attempt to analyse the convergence rate of \ac{TEL} in a practical context was addressed in \cite{rose2014self}. However, the analysis provided in this paper provides a better approximation (as one will see in the sequel). In addition, no attempt has been made to analyse the convergence properties of \ac{ODL}. Third, with the numerical results, we derive the convergence properties of each algorithm. Last, this allows us to provide a comparison between these two algorithms. To the best of our knowledge, this comparison has not been addressed under a practical system model before. 
 
 This paper is organized as follows. Section \ref{sec:models} presents the system model along with a brief description of \ac{TEL} and \ac{ODL}.  Section \ref{sec:results} summarizes the main results of this work. The detailed analysis of the convergence  (i.e. mean convergence  time to a desired state and mean time spent in that state), including the reduction of the \ac{MC}s, is provided in sections \ref{sec:metric_computation},  \ref{sec:Makovchaindimensionnalityreduction}, \ref{sec:approximationMC}, and \ref{sec:Algorithmprocedure}. Numerical results are provided in Section \ref{sec:numerical_results} and Section \ref{sec:conclusion} concludes the paper.


\section{Model}
\label{sec:models}
We consider a network/set of $K$ players $\Players= \{1,\dots, K\}$, that interact among  each other. The players share a set of resources $\Channels=\{r_1,\dots,r_N\}$. Each player $k$ choose an action $a_k$, which consists of  selecting without exchanging any information with the other players a resource inside the set $\Channels$. The vector $\ba=(a_1,\dots,a_K) \in \Actionset$ represents the system action, where $\Actionset=\Channels^K$. The utility received by each player $k\in \Players$ is $u_k(\ba)$, and $\bu(\ba)=(u_1(\ba),\dots,u_k(\ba))$ is the system vector utility.  When two players choose the same resource they will interference with each other. We assume that the utility can take binary values (i.e. $\bu\in\{0,1\}^K$).  Note that, such an hard threshold utility model is commonly encountered in the literature \cite{sheng2014utility,rose2012startTE,simsek2012qlearning}. This problem can be modeled as a normal form game $\mathcal{G}=(\Players,\Actionset,\{u_k\}_{k\in\Players})$. A common approach to solve the aforementioned problem is to study the \ac{PNE} that can be defined as follows.

\begin{definition}[\ac{PNE}]
	An action profile $\ba^*\in \Actionset$ is a \ac{PNE} of game $\Game$ if $\forall k \in \Players$ and $\forall a_k \in \Channels$, $$u_k(a^*_k,a^*_{-k})\geq u_k(a_k,a^*_{-k})$$
\end{definition}
Since we consider a general game model, we make in the following some assumptions in order to ensure the existence of a \ac{PNE}. We suppose that the number of available resources $N$ is greater or equal to the number of players $K$. Furthermore, we assume that if two players interfere with each other (i.e. choose the same resource) then their utilities are equal to 0. The utility of a player is then equal to 1 when no other player choose the same resource. These simplified assumptions can be justified by the fact   that our objective in this paper is to study the performance of \ac{TEL} and \ac{ODL} algorithms and not to study the existence of \ac{PNE} for some game models. It is worth mentioning that even under the above assumptions the problem is still challenging due to the fact that the players cannot communicate with each other and then cannot be aware of the others' actions and they can only observe the result of their own actions (e.g. a player cannot know how many players have chosen the same resource). The resulting Markov chain, as one will see in the sequel, is very complex to analyze under this model.   

In order to deal with the aforementioned decentralized resource allocation problem, two fully distributed  learning schemes, namely \ac{TEL} and \ac{ODL}, can be employed. They have received increasing attention recently, which leads us to analyze their performance and make a comparison between them in this paper. In the remaining of this section, a description of these algorithms is provided. Both algorithms share common characteristics. Each player $k\in \Players$ implements a controller composed with states called moods and noted $m_k$ and, $\bm=(m_1,\dots,m_K)$ is the mood vector of the network. In \ac{TEL}, there are four moods called Content (C), Watchful (W), Hopeful (H) and Discontent (D), whereas  \ac{ODL} controller is solely  composed with the two moods C and D. Furthermore, each player has a benchmark action and a benchmark utility denoted respectively by $\bar{a}_k$ and $\bar{u}_k$. The benchmarks of the network are then denoted by $\bba=(\bar{a}_1,\dots,\bar{a}_K)$ and $\bbu=(\bar{u}_1,\dots,\bar{u}_K)$. At each iteration, every player either selects to use the benchmark action   (i.e. $a_k=\bar{a}_k$) or decides to try a new one $a_k\neq \bar{a}_k$. Then, the player observes the obtained utility $u_k$ and compares to its benchmark utility $\bar{u}_k$. Detailed descriptions of both algorithms, including  the rules used to define/update  of the benchmark actions and utilities,  are provided in the next subsections.


\subsection{\ac{TEL}} 
\label{subsec:TELdesciption}
This section described the rules applied in the \ac{TEL} controller from \cite{pradelski2012learning} of any $k \in \Players$ :
\begin{itemize}
	\item $m_k = C$, there are two cases to consider :
    \subitem 1) with probability $1-\epsilon$, the player keeps playing its benchmark (i.e. $a_k = \bar{a}_k$). The next state changes to H if $u_k>\bar{u}_k$ or, it changes to W if $u_k<\bar{u}_k$ or, it remains C if $u_k=\bar{u}_k$.
    \subitem 2) with probability $\epsilon$, the player experiments a new action, i.e. $a_k \in \Channels \backslash \{\bar{a}_k\}$. The action experimented is selected randomly among $\Channels \backslash \{\bar{a}_k\}$ (i.e. $\Prob{a_k=r_i}=\frac{1}{N-1 }$, $\forall r_i \neq \bar{a}_k$) and, the next state remains $m_k=C$. When $u_k>\bar{u}_k$, player $k$ updates its benchmark with probability $\epsilon^{G(u_k-\bar{u}_k)}$, where $G(x)=-\nu_1 x +\nu_2$, with $\nu_1>0$ and $\nu_2$ such that $0<G(u_k-\bar{u}_k)<1/2$. An update consists in changing the benchmark by the played action and the received utility in the next iteration as follows, $\bar{u}_k\leftarrow{u}_k$ and $\bar{a}_k\leftarrow{a}_k$.
    
	\item $m_k=H$: $a_k=\bar{a}_k$ and the next state changes to C with a utility benchmark update ( i.e. $\bar{u}_k\leftarrow{u}_k$) if $u_k>\bar{u}_k$ or, it changes to W if $u_k<\bar{u}_k$ or, it changes to C if $u_k=\bar{u}_k$.
	\item $m_k=W$: $a_k=\bar{a}_k$ and the next state changes to H if $u_k>\bar{u}_k$ or, it changes to D if $u_k<\bar{u}_k$ or, it changes to C if $u_k=\bar{u}_k$.
	\item $m_k=D$: an action $a_k$ is randomly selected among $\Channels$ (i.e. $\Prob{a_k=r_i}=\frac{1}{N}$, $\forall r_i \in \Channels$) with probability 1. The next state $m_k$ changes to C with probability $\epsilon^{F(u_k)}$, where $F(u)=-\phi_1 u +\phi_2$ with, $\phi_1>0$ and $\phi_2$ such that $0<F(u)<1/{2K}$, with a benchmark update (i.e. $\bar{u}_k\leftarrow{u}_k$ and $\bar{a}_k\leftarrow{a}_k$), otherwise, with probability $1-\epsilon^{F(u_k)}$, $m_k=D$.
\end{itemize}
\subsection{\ac{ODL}} 
\label{subsec:ODLdesciption}
This section described the rules applied in the \ac{ODL} controller from \cite{marden2014achieving} of any player $k \in \Players$ :
\begin{itemize}
	\item $m_k=C$, there are two cases to consider :
    \subitem 1) with probability $1-\epsilon^c$, where $c>K$ is a real constant, $a_k = \bar{a}_k$. If $u_k\neq \bar{u}_k$ then the state $m_k$ changes to D with probability $1-\epsilon^{1-u_k}$. Otherwise, with probability $\epsilon^{1-u_k}$, the cluster updates its benchmark  (i.e. $\bar{u}_k\leftarrow{u}_k$ ) and remains C.
    \subitem 2) with probability $\epsilon^c>0$, a new action is experimented, $a_k \in\Channels \backslash \{\bar{a}_k\}$. The new action is selected randomly in the set $\Channels \backslash \{\bar{a}_k\}$. If  $u_k\neq \bar{u}_k$, the state $m_k$ changes to D with probability $1-\epsilon^{1-u_k}$. Otherwise, with probability $\epsilon^{1-u_k}$, the cluster updates its benchmark (i.e. $\bar{u}_k\leftarrow{u}_k$ and $\bar{a}_k\leftarrow{a}_k$) and remains in C.
    \item $m_k=D$: an action $a_k$ is randomly chosen among $\Channels$. The cluster switches to C with probability $\epsilon^{1-u_k}$ and updates its benchmark (i.e. $\bar{u}_k\leftarrow{u}_k$ and $\bar{a}_k\leftarrow{a}_k$), otherwise with probability $1-\epsilon^{1-u_k}$, it remains D. 
\end{itemize}

 
\subsection{Markov chain representation and performance metrics}
The different states taken by the network are defined by  $\bz=(\bm,\ba,\bba,\bu,\bbu)$ and represent a Markov chain $\Xi_{TEL}$ if the \ac{TEL} is used by all players or $\Xi_{ODL}$ if it is \ac{ODL}. Unless there is an ambiguity, we drop the indices and call the Markov chain $\Xi$. 

The convergence performance of \ac{TEL} and \ac{ODL} is evaluated along two features: {\it i}) the  mean time duration to reach the state maximizing the social welfare, starting from a specific initialization point, also known as \ac{EFHT} and denoted by $T_{EFHT}$, {\it ii}) the mean fraction of time duration spent on that state denoted by $\alpha$. 

It is of interest to note that these algorithms are known to converge under the  {\it interdependence} property (see \cite{pradelski2012learning,marden2014achieving} for the exact definition). In few words, the  {\it interdependence} is the property that for any set of players, there exists an action that changes the utility of a player not in the set. This condition is a sufficient condition as the analysis in \cite{pradelski2012learning,marden2014achieving} was done for more general game model than the one considered in this paper. In our case, the above condition is not needed. In fact, thanks to the presence of the small probability $\epsilon$ in \ac{TEL} and \ac{ODL} (see sections \ref{subsec:TELdesciption} and \ref{subsec:ODLdesciption}), all states of the Markov Chain $\Xi$ communicate and form a unique communication class. $\Xi$ is then ergodic and possesses a unique invariant distribution. This property ensures a non null transition probability between all states for a sufficient number of transitions and a non null probability of the corresponding state. In this paper, we are interested in computing the mean time the system stays in a desired state and the mean time needed to achieve that state for the first time. The desired is the state that maximizes the social welfare of the players. 


\section{Results}
\label{sec:results}
The main result of this work is to provide an efficient approximation of the \ac{MC} for \ac{TEL} and \ac{ODL} algorithms that allow an accurate numerical convergence analysis. The approximated \ac{MC} is denoted by $\widetilde{\Xi}$.  In next sections, we describe the procedure to approximate and reduce the \ac{MC} dimensionality so as to realize the convergence analysis. 
It is worth mentioning that the number of states in the original  \ac{MC} is huge, which makes very hard the computation (even numerically) of the performance metrics  $T_{EFHT}$ and  $\alpha$ for both algorithms. 

Using the proposed efficient approximation, we were able to find interesting results (that are presented in  \ref{sec:numerical_results}). Based on the obtained results,  the following observations  can be highlighted.  We use the Landau notation $\mathcal{O}(.)$ to specify the rate of convergence when $K$ becomes important or when $\epsilon$ is close to 0 but strictly positive. In this notation, $K$ and $\epsilon$ are dropped for clarity.


\begin{observation}
    \label{prop:TELprop}
    For the \ac{TEL}, the \ac{EFHT} $T_{EFHT}=\mathcal{O}(\frac{1 }{\epsilon^{a_1}})$ and $T_{EFHT}=\mathcal{O}(K^{a_2})$ where $a_1,a_2>0$ and, $1-\alpha=\mathcal{O}( \epsilon^{a_3})$ and $1-\alpha=\mathcal{O}( K^{a_4})$ where $a_3,a_4>0$.
\end{observation}

\begin{observation}
    \label{prop:ODLprop}
    For the \ac{ODL}, $T_{EFHT}=\mathcal{O}(\frac{1}{\epsilon^{c b_1}})$ and $T_{EFHT}=\mathcal{O}(b_2^K)$ where $b_1>0$ ,$b_2>1$ and, the stability is $1-\alpha=\mathcal{O}(\epsilon^{ b_3})$ and $\alpha=\mathcal{O}(b_4^K)$,  where $b_3>0$, $1>b_4>0$.
\end{observation}

From these observations some interesting comparisons can be deduced. Both algorithms have a convergence time inversely proportional to $\epsilon$ and a stability that decreases polynomially with $\epsilon$. However, \ac{ODL} has a convergence time which is exponential with respect to $K$ contrary to \ac{TEL} which is polynomial. At low $K$, the convergence time of \ac{ODL} is relatively similar to the \ac{TEL} one, but at higher $K$, \ac{TEL} converges faster than \ac{ODL}. In addition, for \ac{ODL}, the stability decreases exponentially with respect to $K$ whereas, for the \ac{TEL}, the stability decreases polynomially. It follows that the \ac{TEL} is much more stable than the \ac{ODL}.  At low number of players, the convergence time of both algorithms are similar but the stability of \ac{TEL} is better. At higher number of players, the \ac{TEL} performs better than \ac{ODL} for both convergence metrics. These observations result from the analysis of numerical figures of merit computed using the formulas presented in next section.

\section{Metrics computation}
\label{sec:metric_computation}
In this section, we present how to compute the figure of merit of both algorithms using the transition matrix $\bP_0$ of $\Xi$. The method is based on the generalized fundamental matrix $\bF$ for ergodic \ac{MC} developed in \cite{kemeny1981gnrlFund}. The matrix $\bF$, which is an extension of the fundamental matrix introduced in \cite{kemeny1960finite}, is defined by
\begin{equation}
\bF:=(\bI-\bP_0+\bOne \bb^t)^{-1},
\end{equation}
where $\bI$ is the identity matrix, $\bOne$ is a column vector filled with 1, and $\bb$ is any arbitrary column vector such that $\bb^t \bOne \neq 0$. During simulations, we use $\bb=\bOne$.

The first feature concerns the expected first hitting time to a given state $j$ from a state $i$, $\E{T_j}{i}$ and is given by (\cite{kemeny1981gnrlFund} equation (30))
\begin{equation}
\E{T_j}{i}=\frac{\bF_{jj}-\bF_{ij}}{\bpi_j},
\label{eq:FEHT}
\end{equation}
where $\bF_{ij}$ is the term in line $i$ and column $j$ of matrix $\bF$, and $\bpi_j$ is the stationary probability of state $j$. The stationary  distribution is given by the property that (equation (28) from~\cite{kemeny1981gnrlFund})
\begin{equation}
\bb^t \bF = \bpi.
\label{eq:stationrydistribution}
\end{equation}
The second feature that describes the performance of stochastic stable algorithms is the mean fraction of time spent in the state that maximizes the social welfare. In an ergodic \ac{MC}, the proportion of time $\alpha_j$ spent in a state $j$ is equal to its stationary probability $\alpha_j=\bpi_j$ (\cite{kemeny1960finite} Theorem 4.2.1) that can be computed using \eqref{eq:stationrydistribution}.

The convergence analysis realization requires the manipulation of transitions matrices. The huge number of states $\cS$ grows exponentially with $K$ and $N$ (see Section~\ref{sec:complexity}) and since $\Xi$'s transition matrix has dimension $(\cS\times\cS)$, it needs to be approximated to allow numerical computation of the performance. As an example, even for small values $N=K=3$, the number of states is already $\cS=373,248$. In this work, we propose a new approach to build the approximated Markov chain $\widetilde{\Xi}$ whose transition matrix is noted $\bP$. Note that the approximation $\widetilde{\Xi}$ is built such that it is ergodic like $\Xi$ which means that, $\bP$ admits a unique invariant distribution with strictly positive components. With this approximation, formulas \eqref{eq:FEHT} and \eqref{eq:stationrydistribution} are still valid if $\bP_0$ is replaced by $\bP$. It remains to construct $\bP$ with justified and motivated arguments. This approach follows two steps: {\it i}) first we approximate the original Markov chain $\Xi$ by identifying some invariance induced by the utility model, {\it ii}) we then further reduce the Markov chain complexity by neglecting some transitions. 
Then, from the probability transition matrix of the approximated Markov chain, we are able to compute the two convergence figures of merit.


\section{Reducing the Markov chain dimensionality}
\label{sec:Makovchaindimensionnalityreduction}
In order to approximate $\Xi$, we start by  considering only the states called {\it recurrence classes of the unperturbed process}  \cite{young1993evolution}, shorten as  \ac{RC}, that were used as the key feature for the \ac{TEL} and \ac{ODL} proof of convergence. The system tends to spend naturally a high amount of time in those states which thus play a major role in convergence metrics. The reason comes from the combination of two properties. First, the network needs at least one experimentation to leave an \ac{RC}, which occurs with small probability $\epsilon$. Secondly, by definition, the network always naturally goes to an \ac{RC} when no perturbation occurs. These states are characterized by $\bm=\bm_C:=(C,C,\dots,C)$, $\ba=\bba$, and $\bu=\bbu$, i.e., all the players are in the content mood and aligned (i.e. $\ba=\bba$, and $\bu=\bbu$). We denote by $\cR$ the set of these states. We can also drop some notations, and we rewrite a state $\bz=(\bm_C,\ba,\bba,\bu,\bbu)\in \cR$ as $\bz=(\bba,\bbu)$

To reduce the number of \ac{RC}, two invariances induced by the utility model are highlighted. First of all, due to the binary utility values and the utility rules, interchanging actions between players is equivalent to interchange the utility vector components accordingly, thus not modifying the number of 0 and 1 in the utility vector. As such, we can deduce that it does not change the ``global'' performance of the network.  For instance, let consider a network with three players and three resources $\bz_1 =((r_1,r_2,r_1),(0,1,0))\in \cR$. Players 1 and 3 have null utility because they use the same resource. If we interchange the actions of player 1 and 2, $\bz_1$ is transformed into $\bz_2=((r_2,r_1,r_1),(1,0,0))$. There is always one player with utility 1 and two players with utility 0. Nothing has changed from a network perspective, thus, the algorithm performance remains the same from these two states.
 
Secondly, notice also that interchanging the resource labels does not change at all the utility vector (this is also true for geographical models when orthogonality between frequencies is assumed). For instance, if we change the resource label 1 with label 3 and, label 2 with label 1, $\bz_1$ becomes $\bz_3=((r_3,r_1,r_3),(0,1,0))$ which involves the same observations as the previous modification. 

These two invariances have led us to represent any \ac{RC} with the ordered repartition of players over resources. For any action vector $\bba$, we build the repartition vector of players over resources $\bd:=(d_{1},d_{2},\cdots,d_{N})$ where $d_{i}=\sum_{k=1}^{K}\ind{a_k=f_i}$ is the number of players that use a resource $f_i\in \Channels$. For instance, the repartition of player in $\bz_1$ is $\bd_1=(2,1,0)$ and the repartition vector of $\bz_3$ is $\bd_3=(1,0,2)$. The ordered repartition vector
is $\bs=(s_1,\dots,s_N)$ where $\forall i,j \in [1,N]$, $i<j$, $\exists i',j' \in [1,N]$, $s_i=d_{i'}\geq s_j=d_{j'}$. For instance, the ordered repartition vector of $\bz_1$ is $\bs_1=(2,1,0)$ and, the ordered repartition of $\bz_3$ is $\bs_3=(2,1,0)$ which is equal to $\bs_1$. Thus, it follows that this representation makes no difference between \ac{RC} that are invariant with respect to the transformations mentioned. Hence, it is possible to reduce the number of \ac{RC} in $\Xi$. Moreover, the utility vector repartition is directly specified by the ordered repartition of players, then we drop this notation and $\bz=(\bba,\bbu)$ becomes $\bz=\bs$.

 In what follows, for ease of comprehension we slightly modify  \ac{RC} notations. For each $\bz\in \cR$ the number of resources employed is noted $n=\sum_{\ell=1}^{N} \ind{s_\ell > 0 }$. In addition, for each $n \in [1,N]$ there exists different possible ordered repartitions of players whose number is noted  $\text{I}_N(n)$. It is equal to the number of ways to partition integer $N$ in $n$ parts, i.e. $\text{I}_N(n)=\text{Part}(N,n)$ where the recursive formula gives $\text{Part}(N,n)=\text{Part}(N-1,n-1)+\text{Part}(N-n,n),$
 and for any integers $x$, $y$, $\text{Part}(x,x)=1$,  $\text{Part}(x<y,y)=0$ and $\text{Part}(x,1)=1$ (\cite{comtet1974partitions} Chapter 2, Section 2.1, Theorem B).
Thus, any $\bz \in \cR$ can be noted $\bZ_n(i)$ where $n$ is the number of resources used and $i\in [1,\text{I}_N(n)]$ is the indices of the ordered repartition and, the associated ordered repartition vector is $\bS_n(i)=(S_{n,1}^i,S_{n,2}^i,\dots,S_{n,N}^i)$. For instance, in a network with $N=K=4$, when $n=2$ there are two possible ordered repartition $\bS_2(1)=(3,1,0,0)$ and $\bS_2(2)=(2,2,0,0)$. However, for $n=3$ there is a unique repartition $\bS_3(1)=(2,1,1,0)$. The mapping between indices $i$ and the ordered repartitions is arbitrary and has to be made by the experimenter. The reduced states $\bZ_n(i)$ for all $n\in [1,N]$ and for all $i\in [1,I_N(n)]$ are called \ac{RRC}.

Notice that the social welfare of \ac{RC} represented by the same \ac{RRC} are equal, but we can find different \ac{RRC} for which their elements have the same social welfare.

\section{Approximated Markov chain}
\label{sec:approximationMC}

In this section, we build an approximation $\widetilde{\Xi}$ of $\Xi$ that is composed of the \ac{RRC} and a subset of intermediary states between \ac{RRC}. More specifically, the construction of the intermediary states considered in each approximations (\ac{TEL} and \ac{ODL}) is detailed. These constructions are driven by, {\it i)} the willingness to conserve the ergodic property of $\Xi$ in order to be able to approach its convergence performance, {\it ii)} the need to construct a Markov chain with low dimension (i.e. with the least number of states). A condition to make property {\it i)} realizable, consists in constructing intermediary states around each \ac{RRC} such that, all states of $\widetilde{\Xi}$ (i.e. \ac{RRC} and intermediary states) are accessible from one another. We note $\xi^n(i)$ the set that contains the \ac{RRC} $\bZ_n(i)$ and some associated intermediary states that we define later. The simplest, thus verifying {\it ii)}, and necessary way to conserve the ergodicity property is to construct intermediary states such that, if the transition between sets $\xi^{n}(i) \leadsto \xi^{n+1}(j)$ exists then there also exists a transition from sets $\xi^{n+1}(j) \leadsto \xi^{n}(i)$ (the symbol $\leadsto$ specifies that this transition can involve multiple states in $\widetilde{\Xi}$). This is the consequence of the fact that from every \ac{RRC} $\bZ_n(i)$ where $n<N$, there are players interfered and, it is possible for one of them to find a free resource (e.g. an interfered player experiments on a free resource). Repeating this process successively shows that, all \ac{RRC} can access $\bZ_N(1)$, which is the \ac{RRC} without interference. Therefore, the condition, if  $\xi^{n}(i) \leadsto \xi^{n+1}(j)$ exists, then, so does $\xi^{n+1}(j) \leadsto \xi^{n}(i)$, implies that all sets communicate. Finally, the previous condition becomes sufficient, if the sets are constructed such that all states in all sets are accessible. In addition to these simplifications, we consider the following hypothesis to build $\widetilde{\Xi}$ completely. 
\begin{ass}
    \label{ass:ass1}
    {\it For each algorithm models, we assume at each iteration of the algorithm that at most one content player can experiment, and such, solely when the system is in an all content mood and aligned state, i.e. $\bm=\bm_C$, $\bu=\bar{\bu}$ and, $\ba=\bar{\ba}$.}
\end{ass}
The reason to propose this assumption is summarized as follows. When all players are content, the probability that one player experiments (i.e. $0<\epsilon\ll 1$) is larger than the probability that two or more player experiment (i.e $0<\epsilon^2\ll\epsilon\ll 1$). Moreover, when the system is not aligned, it goes in less than two steps and with an high probability (i.e. $\approx (1-\epsilon)^2$) to a state in which all players are content and aligned or, that contains a discontent player. Thus, most of the time, the system is either in {\it a)} an all content and aligned state or, {\it b)} it contains at least one discontent player. In case {\it a)}, it is most probable that only one player experiments whereas, in case {\it b)}, the probability that a discontent player experiments is 1 which is much more important than the probability for a content player to experiment ($\epsilon \ll 1$).
\begin{hypothesis}
    \label{hyp:hyp1}
    {\it For \ac{TEL} model, the probability that a discontent player accepts a new utility u as a benchmark is $1 \geq\epsilon^{F(u)}\geq \epsilon^{\frac{1}{2K}}$. We suppose that $\epsilon^{F(0)}=\epsilon^{\frac{1}{2K}}$ and that, $\epsilon^{F(1)}=\epsilon^{0}=1$.}
\end{hypothesis}
In other words, we suppose that the constants $\phi_1$ and $\phi_2$ from section \ref{sec:models} have been chosen such that $F(.)$ spans the whole available region.

In next two sections, we present the constructions of sets $\xi^n(i)$ of each algorithm. We start the reasoning by considering all sets $\xi^n(i)=\{\bZ_n(i)\}$. Then we add successively intermediary states in all sets to build the approximated Markov chain. When a state is added to $\xi^n(i)$ it is also added to any other set $\xi^{n'}(i')$ where $i\neq i'$ and $n\neq n'$. Figures \ref{fig:MC_TE} and \ref{fig:MCmodODL} present, for ease of space and comprehension, a resulting  partial view of $\widetilde{\Xi}$'s intermediary models with two sets $\xi^n(i)$ and $\xi^{n+1}(j)$ for \ac{TEL} and \ac{ODL} respectively. The lines define the oriented connections between states. Plain lines correspond to direct transition inside the same set $\xi^n(i)$ whereas dashed lines correspond to direct transitions between different sets. The connexions are detailed in appendix \ref{app:probTEL} and \ref{app:probODL} for the \ac{TEL} and \ac{ODL} respectively. In these figures, without loss of generality, it is supposed that, there exists $j$ such that $\xi^n(i)$ is connected to $\xi^{n+1}(j)$. In such a case, we would also like to have $\xi^{n+1}(j)$ connected to $\xi^n(i)$ for ergodicity. We also suppose that, all intermediary states are present for simplicity of comprehension, whereas as explained in next two sections, there exists some conditions in which they have to be deleted from their corresponding set to keep $\widetilde{\Xi}$ ergodic.
\subsection{\ac{TEL} model}
\label{subsec:TELmodel}
This section presents the construction of the intermediary states in the approximated Markov chain based on the \ac{TEL} algorithm described in section \ref{subsec:TELdesciption}. Given any \ac{RRC} $\bZ_n(i)$, a transition where a player interfered finds a free resource, e.g. $\bZ_{n}(i) \leadsto \bZ_{n+1}(j)$, does not necessitate additional intermediary state unless one player is left alone on its resource after the experimentation. In this situation, the left alone player sees its utility increases and becomes hopeful. Therefore, we start by considering in $\xi^n(i)$ the state $\xi_0^n(i)$ in addition to $\bZ_n(i)$ where 
\begin{itemize}
    \item $\xi_0^n(i)$ corresponds to a player alone in $\bZ_n(i)$ that is hopeful.
\end{itemize}
Thus, at this step, $\forall n, i$, $\xi^n(i)=\{\bZ_n(i),\xi_0^n(i)\}$.

 A transition in which the network uses one less frequency, e.g. $\xi^{n+1}(j) \leadsto\xi^{n}(i)$, involves a player that accepts a lower benchmark, which is only possible through a discontent mood. To become discontent, a player passes through a watchful mood. This leads us to consider the two intermediary states $\xi_1^{n}(i)$ and $\xi_2^{n}(i)$ where
\begin{itemize}
    \item $\xi_1^n(i)$ is the state where a player alone in $\bZ_n(i)$ is watchful, 
    \item $\xi_2^n(i)$ is the state where a player alone in $\bZ_n(i)$ is discontent. It corresponds to the situation where the watchful player in $\xi_1^n(i)$ experiences one more iteration a decrease in utility.
\end{itemize}
Note that during the transition $\xi_1^n(i) \rightarrow \xi_2^n(i)$ (where $\rightarrow$ means that the transition is direct), the system is not aligned whereas, a content player experiments. It is not in accordance with hypothesis 1 but, this is the \textbf{only time} that the hypothesis 1 is overrided in order to keep the chain ergodic. Finally, to avoid any absorbing state two more intermediary states $\xi_3^{n}(i)$ and $\xi_4^{n}(i)$ are considered where
\begin{itemize}
    \item $\xi_3^n(i)$ is a state where two players that were alone in $\bZ_n(i)$ are using the same resource and one of them is watchful. It corresponds to the case where the discontent player from $\xi_2^n(i)$ has updated its benchmark with the resource of a player that was not interfered in $\bZ_n(i)$.
    \item $\xi_4^n(i)$ is a state where two players that were alone in $\bZ_n(i)$ are using the same resource and one of them is discontent. It corresponds to the state that follows $\xi_3^n(i)$ where the player watchful becomes discontent.
\end{itemize}

The base to construct our model for \ac{TEL} is established with $\xi^n(i)=\{\bZ_n(i),\xi_0^n(i), \xi_1^n(i),$ $\xi_2^n(i), \xi_3^n(i), \xi_4^n(i)\}$.  It is said in the introduction, that all intermediary states have to be accessible but, in some cases they are not all present. For instance, when every player in $\bZ_n(i)$ is interfered, no one can become discontent and states $\xi_1^n(i)$, $\xi_2^n(i)$, $\xi_3^n(i)$ and $\xi_4^n(i)$ are not present. These absences have to be taken into account, to compute the probabilities in appendix \ref{app:probTEL}, and during simulations in order to build an ergodic chain (an isolated state in a matrix makes the chain not ergodic). These cases are described as follows starting with any given $\xi^n(i)=\{\bZ_n(i)\}$:
\begin{itemize}
    \item If in $\bZ_n(i)$ all players are interfered, only the state $\bZ_n(i)$ is present
    \item If in $\bZ_n(i)$ one player is alone on its resource, this player can become discontent or hopeful, however, it cannot make an other player discontent. Therefore, include states $\xi_0^n(i),\xi_1^n(i)$ and $\xi_2^n(i)$ in $\xi^n(i)$. There is one exception, where the distribution $\bS_n(i)$ is of the form $(2,\dots,2,1,0,\dots,0)$ and, the state $\xi_0^n(i)$ is removed from $\xi^n(i)$.
    \item If in $\bZ_n(i)$ at least two players are alone on their respective resource, include $\xi_3^n(i)$ and $\xi_4^n(i)$ in $\xi^n(i)$.
\end{itemize}

The transitions between states and the associated probabilities are detailed in appendix \ref{app:probTEL}.
\subsection{\ac{ODL} model}
\label{subsec:ODLmodel}
This section presents the construction of the intermediary states in the Markov chain approximation based on the \ac{ODL} algorithm described in section \ref{subsec:ODLdesciption}. First of all, the model that contains only the $\xi^n(i)=\{\bZ_n(i)\}$ is sufficient to have an ergodic chain $\widetilde{\Xi}$. The transition $\bZ_{n}(i) \rightarrow \bZ_{n}(i)$ occurs if nothing happens. The transition $\bZ_{n}(i) \rightarrow \bZ_{n+1}(j)$ represents an interfered player that experiments and finds a free resource. The reversed transition $\bZ_{n+1}(i) \rightarrow \bZ_{n}(j)$ occurs if one of the not interfered player in $\bZ_{n+1}(j)$ goes back to the position of the experimenter from $\bZ_{n}(i)$. The accuracy of the model can be increased by adding a few more states. The stability of \ac{ODL} is directly related to the number of discontent players. Such players experiment randomly, which makes the number of possible transitions between states growing very fast with the number of discontent players. It prevents us from describing too many discontent players at the same iteration. In this work, we manage to model the case where at most two players can be discontent at the same time. Going beyond this would require for each additional discontent player a large amount of  extra transitions for a small accuracy gain. This model requires three more states $\xi_1^n(i)$, $\xi_2^n(i)$ and $\xi_3^n(i)$ to be added with each \ac{RRC} $\bZ_n(i)$:
\begin{itemize}
    \item $\xi_1^n(i)$ corresponds to the case where a player alone in $\bZ_n(i)$ is discontent, 
    \item $\xi_2^n(i)$ corresponds to the case where two players alone in $\bZ_n(i)$ are discontent.
    \item $\xi_3^n(i)$ is a state where one of two players that share the same resource in $\bZ_n(i)$ is discontent.
\end{itemize}

Like in previous section \ref{subsec:TELmodel}, there are some cases, depending on $\bZ_n(i)$, where $\xi_1^n(i)$,  $\xi_2^n(i)$ and $\xi_3^n(i)$ are not all present simultaneously in $\xi^n(i)$. They have to be removed accordingly to make the resulting Markov chain ergodic. These cases are described as follows starting with  $\xi^n(i)=\{\bZ_n(i)\}$:
\begin{itemize}
    \item If there exists a resource played by two players in $\bZ_n(i)$, include the state $\xi_3^n(i)$ in the set $\xi^n(i)$.
    \item If at least one player in $\bZ_n(i)$ is alone on its resource, include the state $\xi_1^n(i)$ in $\xi^n(i)$.
    \item If at least two players in $\bZ_n(i)$ are alone on their respective resource, include $\xi_2^n(i)$ in $\xi^n(i)$.
\end{itemize}

The transitions between states and the associated probabilities are detailed in appendix \ref{app:probODL}.

\begin{figure}[t!]
    \centering
    \subfloat[\ac{TEL}.]{
    \includegraphics[width=0.55\textwidth]{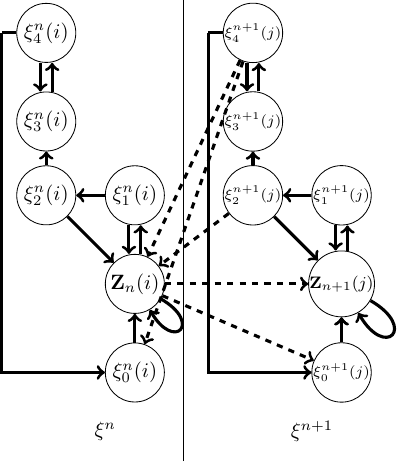}
    \label{fig:MC_TE}}
    \subfloat[\ac{ODL}.]{
    \includegraphics[width=0.35\textwidth]{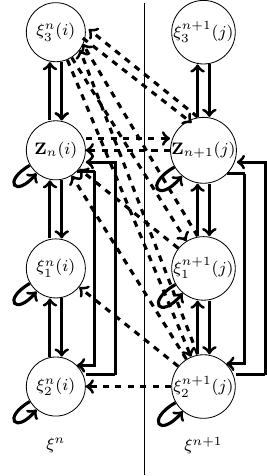}
    \label{fig:MCmodODL}}
    \caption{Partial view of $\widetilde{\Xi}$ for both algorithms.}
    \vspace{-0.75cm}
\end{figure}


\subsection{Complexity comparison}
\label{sec:complexity}
\begin{figure}[t!]
    \centering
    \includegraphics[width=0.4\textwidth]{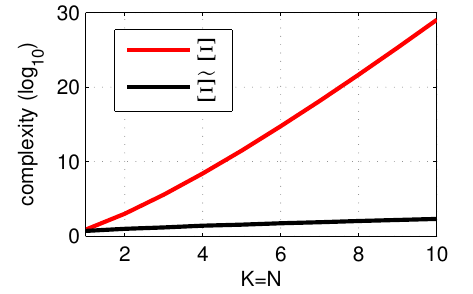}
    \caption{Complexity comparison between $\Xi$ and $\widetilde{\Xi}$.}
    \label{fig:complexity}
    \vspace{-0.75cm}
\end{figure}

We compute the Markov chain complexities, to highlight the importance of the transformations from $\Xi$ to $\widetilde{\Xi}$ made in this work. The simplifications and approximations are essential in order to be able to predict the algorithm performance. The number of states in $\Xi$ is given by the product of component dimension of $\mathbf{z}=(\bm,\ba,\bba,\bu,\bbu)$. The vector of player moods can have $M^K$ values if the mood of each player can take $M$ values. The vector of player actions and action benchmarks $\ba$ or $\bba$ can take $N^K$ values each one. The utility vector $\bu$ is specified by the action vector $\ba$ and, the utility benchmark vector $\bbu$ can take $2^K$ values. Therefore, the complexity of $\Xi$ is $(M N ^2 2)^K$. This is obviously intractable and, we have reduced the recurrence states $\cR$ into $\bZ$ which has a cardinality $|\bZ|=\sum_{n=1}^{K} \text{Part}(K,n)$.  Afterwards, we have approximated $\Xi$ by keeping some intermediary states as detailed in previous section \ref{sec:approximationMC}. Figure~\ref{fig:complexity} presents the complexity of $\Xi$ and $\widetilde{\Xi}$ for the \ac{TEL} algorithm with respect to the number of players. The significant complexity reduction allows us to predict performance numerically.

\section{Procedure to compute the transition matrix}
\label{sec:Algorithmprocedure}
Once the states of $\widetilde{\Xi}$ are established, the next step consists in computing the transition probabilities of matrix  $\bP$. The procedure is described in algorithm~\ref{alg:algo} and summarized as follows.   The first step necessitates to generate all \ac{RRC}s. For this purpose, a classical integer partitioning algorithm is used to generate all ordered repartition vector $\bS_n(i)$ \cite{knuth2005art}. The number of \ac{RRC} using $n\in [1,\dots,N]$ resources among $N$ is given by $\text{I}_N(n)$ (see section \ref{sec:Makovchaindimensionnalityreduction}). In both algorithms, at each \ac{RRC} $\bZ_n(i)$ is associated an intermediary state model $\xi^n(i)$, whose number of states depends on some exceptions specified in sections \ref{subsec:TELmodel} and \ref{subsec:ODLmodel} for the \ac{TEL} and \ac{ODL} respectively. One has to pay attention to these exceptions when it makes the one-to-one mapping function between the states of $\widetilde{\Xi}$ and the lines of $\bP$. Then, the algorithm 1 goes trough all $\bZ_n(i)$ and looks for all $j\in I_N(n+1)$ such that $\bZ_{n+1}(j)$ is accessible from  $\bZ_n(i)$. When $n<N$, there exists at least one such a $j$ and, by construction the set $\xi^{n+1}(j)$ is connected to the set $\xi^{n}(i)$. The transition probabilities are computed in the algorithm through three consecutive steps. These steps and formulas are highlighted in the same order in  appendices \ref{app:probTEL} and \ref{app:probODL} for the \ac{TEL} and \ac{ODL} respectively. On the first hand, the probabilities inside the set $\xi^n(i)$ are computed. On the second and third hand, for each $j$ in $I_N(n+1)$ such that $\xi^n(i)$ is connected to $\xi^{n+1}(j)$, the algorithm computes, the probabilities from set $\xi^n(i)$ to set $\xi^{n+1}(j)$ and,  the reverse probabilities from set $\xi^{n+1}(j)$ to set $\xi^n(i)$.

\begin{figure}[t!]
	\centering
	\includegraphics[width=0.6\textwidth]{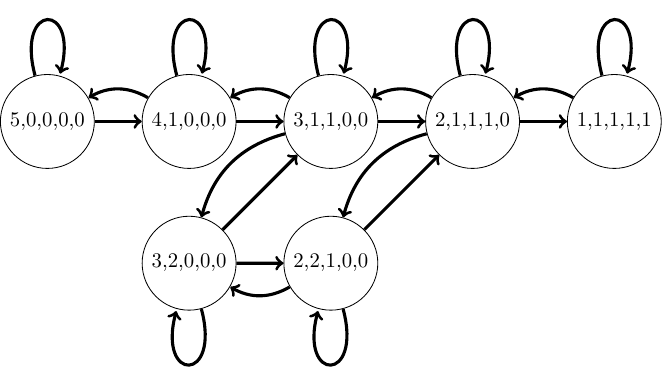}
	\caption{Example of transitions considered between \ac{RRC} in our models for $N=K=5$.}
	\label{fig:MacroStates}
    \vspace{-0.75cm}
\end{figure}

 The example provided in figure~\ref{fig:MacroStates} with $K=N=5$ highlights the links between the sets $\xi^n(i)$, identified by the vector $\bS_n(i)$. For instance, the set in the top left corresponds to 5 players interfering on the same resource.

\begin{algorithm}
	\caption{Computing the matrix $\bP$ - Part 1/2}
	\begin{algorithmic}[1]
		\renewcommand{\algorithmicrequire}{\textbf{Input:}}
		\renewcommand{\algorithmicensure}{\textbf{Output:}}
		\Require $\bZ$; $\forall n\in [1,N]$, $\text{I}_N(n)$; $\forall n\in [1,N]$, $\forall i\in [1,I_N(i)]$ 
		\Ensure $\bP$
        \State Generate $\forall n\in [1,N]$ and $\forall i \in [1,I_N(n)]$, $\bZ_n(i)=(\bS_n(i))$ with an integer partitioning algorithm, and construct each set $\xi^n(i)$ following rules in section \ref{sec:approximationMC}
		\For {$n = 1$ to $N$} 
		\For {$i = 1$ to $\text{I}_N(n)$}
		\State Select a distribution $\mathbf{S}_n(i)=[S_{n,1}^i,S_{n,2}^i,\cdots,S_{n,n}^i,0,0,\cdots,0]$
		\State  Compute the following probabilities using appendices~\ref{app:probTEL} and~\ref{app:probODL} for \ac{TEL} and \ac{ODL} respectively (check the existence of links using exceptions from sections \ref{subsec:TELmodel} and \ref{subsec:ODLmodel}), and fill the matrix $\bP$:
		\begin{itemize}         
            \item(\ac{TEL}) $p_{\bZ_n(i)\xi_1^n(i)}$, $p_{\bZ_n(i)\bZ_n(i)}$, $p_{\xi_1^n(i)\xi_2^n(i)}$, $p_{\xi_1^n(i)\bZ_n(i)}$, $p_{\xi_2^n(i)\bZ_n(i)}$, $p_{\xi_2^n(i)\xi_3^n(i)}$, $p_{\xi_2^n(i) \xi_2^n(i)}$, $p_{\xi_3^n(i)\xi_4^n(i)}$, $p_{\xi_4^n(i)\xi_0^n(i)}$, $p_{\xi_4^n(i)\xi_3^n(i)}$, $p_{\xi_4^n(i)\xi_4^n(i)}$, $p_{\xi_0^n(i)\bZ_n(i)}$, $p_{\xi_0^n(i)\bZ_n(i)}$ using  \eqref{eq:Znitoxi1niTEL}, \eqref{eq:znitozniTEL}, \eqref{eq:xi1nitoxi2niTEL}, \eqref{eq:xi1nitozniTEL}, \eqref{eq:xi2nitoZniTEL}, \eqref{eq:xi2nixi3niTEL}, \eqref{eq:xi2nitoxi2niTEL}, \eqref{eq:xi3nitoxi4niTEL}, \eqref{eq:xi4nitoxi0niTEL}, \eqref{eq:xi4nitoxi3niTEL}, \eqref{eq:xi4nitoxi4niTEL}, \eqref{eq:xi0nitozniTEL} respectively,
            
            \item(\ac{ODL}) $p_{\bZ_n(i)\xi_1^n(i)}$, $p_{\bZ_n(i)\xi_2^n(i)}$, $p_{\bZ_n(i)\xi_3^n(i)}$, $p_{\bZ_n(i)\bZ_n(i)}$, $p_{\xi_1^n(i)\bZ_n(i)}$, $p_{\xi_1^n(i) \xi_2^n(i)}$, $p_{\xi_1^n(i) \xi_1^n(i)}$, $p_{\xi_2^n(i)\bZ_n(i)}$, $p_{\xi_2^n(i)\xi_1^n(i)}$, $p_{\xi_2^n(i)\xi_2^n(i)}$, $p_{\xi_3^n(i) \bZ_n(i)}$, $p_{\xi_3^n(i)\xi_3^n(i)}$ using 
            \eqref{eq:znitoxi1inODL}, \eqref{eq:znitoxi2niODL}, \eqref{eq:znitoxi3niODL}, \eqref{eq:znitozniODL}, \eqref{eq:xi1nitozniODL}, \eqref{eq:xi1nitoxi2niODL}, \eqref{eq:xi1nitoxi1niODL}, \eqref{eq:xi2nitozniODL}, \eqref{eq:xi2nitoxi1niODL}, \eqref{eq:xi2nitoxi2niODL},  \eqref{eq:xi3nitozniODL}, \eqref{eq:xi3nitoxi3niODL}, respectively.
        \end{itemize}
        \algstore{testcont}
    	\end{algorithmic} 
        \label{alg:algo}
    \end{algorithm}
    \begin{algorithm}
        \ContinuedFloat
        \caption{Computing the matrix $\bP$ - Part 2/2}
        \begin{algorithmic}[1]
        \algrestore{testcont}
		\For {$k = 1$ to $n$}
		\If  {$S_{n,k}^i>1$}
		\State $w\leftarrow(S_{n,1}^i,\cdots,S_{n,k}^i-1,\cdots,S_{n,n}^i,1,0,\cdots,0)$
		\State $\tilde{w}\leftarrow w$  sorted in decreasing order
		\State Find $j \in \text{I}_{N}(n+1)$ such that  $\mathbf{S}_{n+1}(j)=\tilde{w}$ which corresponds to state $\bZ_{n+1}(j)$
		\State Compute the following probabilities using appendices~\ref{app:probTEL} and~\ref{app:probODL} for \ac{TEL} and \ac{ODL} respectively (check the existence of links using exceptions from sections \ref{subsec:TELmodel} and \ref{subsec:ODLmodel}), and fill the matrix $\bP$:
		\begin{itemize}
            \item (\ac{TEL})  $p_{\bZ_n(i)\bZ_{n+1}(j)}$, $p_{\bZ_n(i)\xi_0^{n+1}}$,  $p_{\xi_2^{n+1}(j)\bZ_n(i)}$, $p_{\xi_4^{n+1}(j) \bZ_n(i)}$, $p_{\xi_4^{n+1}(j)\xi_0^n(i)}$  using \eqref{eq:znitoznp1jTEL}, \eqref{eq:znitoxi0np1jTEL}, \eqref{eq:xi2np1jtozniTEL}, \eqref{eq:xi4np1jtozniTEL}, \eqref{eq:xi4np1jtoxi0niTEL} respectively,
            \item (\ac{ODL}) $p_{\bZ_n(i)\bZ_{n+1}(j)}$, $p_{\bZ_n(i)\xi_1^{n+1}(j)}$, $p_{\bZ_n(i)\xi_2^{n+1}(j)}$, $p_{\xi_3^n(i)\bZ_{n+1}(j)}$, $p_{\xi_3^n(i)\xi_1^{n+1}(j)}$, $p_{\xi_3^n(i)\xi_2^{n+1}(j)}$, $p_{\bZ_{n+1}(j)\bZ_n(i)}$, $p_{\bZ_{n+1}(j)\xi_3^n(i)}$, $p_{\xi_1^{n+1}(j)\bZ_n(i)}$,  $p_{\xi_1^{n+1}(j)\xi_3^n(i)}$, $p_{\xi_2^{n+1}(j)\bZ_n(i)}$, $p_{\xi_2^{n+1}(j)\xi_1^n(i)}$, $p_{\xi_2^{n+1}(j)\xi_2^n(i)}$ and $p_{\xi_2^{n+1}(j)\xi_3^n(i)}$  using \eqref{eq:ZnitoZnp1jODL}, \eqref{eq:Znitoxi1np1jODL}, \eqref{eq:Znitoxi2np1jODL},\eqref{eq:xi3toZnp1ODL}, \eqref{eq:xi3toxi1np1ODL}, \eqref{eq:xi3toxi2np1ODL}, \eqref{eq:Znp1jToZniODL}, \eqref{eq:Znp1jToxi3niODL}, \eqref{eq:xi1np1jtozniODL}, \eqref{eq:xi1np1jToxi3niODL}, \eqref{eq:xi2np1zniODL}, \eqref{eq:xi2np1jxi1iODL}, \eqref{eq:xi2np1jxi2niODL}, \eqref{eq:xi2np1jxi3niODL} respectively.
        \end{itemize}
		\EndIf
		\EndFor
		\EndFor
		\EndFor\\
		\Return $\bP$ 
	\end{algorithmic} 
\end{algorithm}

\section{Numerical results}
\label{sec:numerical_results}

\subsection{Accuracy of the proposed models}
\label{subsec:num_result_conv_analysis}
We assess the accuracy of our proposed models by comparing the values obtained for $\E{T_j}{i}$ using \eqref{eq:FEHT} and $\alpha_{j}$ using \eqref{eq:stationrydistribution}  of $\widetilde{\Xi}$, with Monte Carlo simulations. We consider three different values for $K=\{3, 5,7\}$ in two cases $N=K$ and, $N=K+2$. In \ac{ODL}, the constant $c$ is equal to $K$. Both algorithms are compared with respect to the same probability to experiment from a content mood, i.e. $\epsilon$ in \ac{TEL} is equal to $\epsilon^{c}$ in \ac{ODL}. The \ac{EFHT} $\tau_{ij}$ is computed from, the state $i$ where all players are on the same resource (e.g. state with $\bS_1(1)=(5,0,0,0,0)$ in Figure \ref{fig:MacroStates}), to the state $j$ where they are all on different resources (e.g. state with $\bS_5(1)=(1,1,1,1,1)$ in Figure \ref{fig:MacroStates}). The stability $\alpha$ is computed for the state $j$ where all players use a different resource. In Monte Carlo simulations with use 5000 trials to compute $\tau_{ij}$ and $10^6$ trials to compute $\alpha$.

\begin{figure}[t!]
    \centering
    \subfloat[\ac{TEL} : convergence time]{\includegraphics[width=0.4\textwidth]{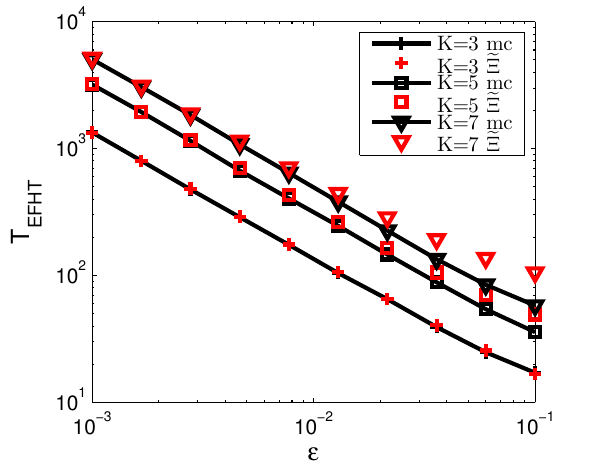}
        \label{fig:TE_FEHT}}\hspace{0.5cm}
    \subfloat[\ac{TEL} : fraction of time ]{\includegraphics[width=0.4\textwidth]{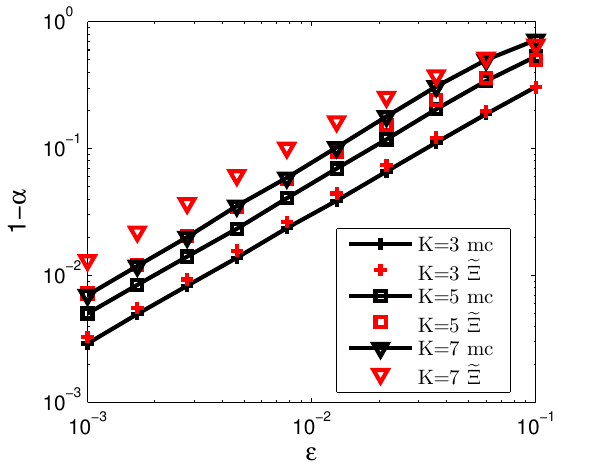}
        \label{fig:TE_proptime}}
    \caption{\ac{EFHT} comparisons between our approximated models and Monte Carlo simulations when $N=K$.}
    \label{fig:FEHT}
\end{figure}

For \ac{TEL} algorithm, Figures \ref{fig:TE_FEHT} and \ref{fig:TE_proptime} present the \ac{EFHT} and the fraction of time $1-\alpha$ when $K=N$ respectively. The reason to display $1-\alpha$ instead of $\alpha$ is to discern the values close to one at low $\epsilon$. For both features, these results are accurate in comparison to Monte Carlo simulations. The \ac{EFHT} converges to the Monte Carlo results when $\epsilon$ decreases. The little gap observed at higher $\epsilon$ is caused by an increasing probability to have more than one experiment at a time. Thus, the probability for the system to not be aligned increases and, Assumption 1 is less valid. The offset observed in Figure \ref{fig:TE_proptime} is due to the fact that, we are able to represent accurately at most one discontent player at each algorithm iteration. The stability is highly related to the number of discontent players.

\begin{figure}[t!]
    \centering
    \subfloat[\ac{TEL} : convergence time]{\includegraphics[width=0.4\textwidth]{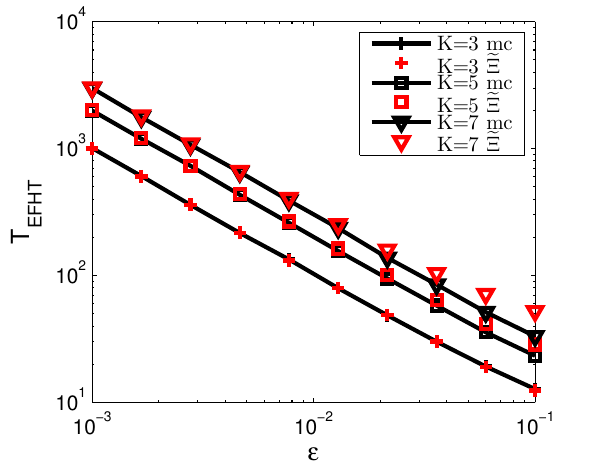}
        \label{fig:TE_FEHT_Cp2K}}\hspace{0.5cm}
    \subfloat[\ac{TEL} : fraction of time]{\includegraphics[width=0.4\textwidth]{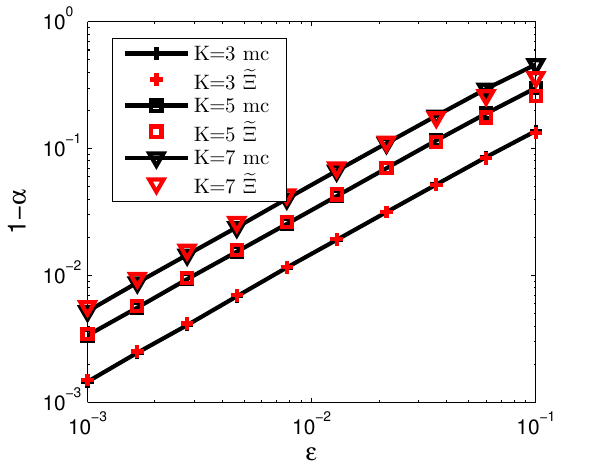}
    \label{fig:TE_proptime_Cp2K}}
    \caption{\ac{EFHT} and stability comparisons between our approximated models and Monte Carlo simulations when $N=K+2$.}
    \label{fig:FEHT_Cp2K}
    \vspace{-0.75cm}
\end{figure}

Figure \ref{fig:TE_FEHT_Cp2K} and~\ref{fig:TE_proptime_Cp2K} present the same results but with $N=K+2$. The goal is to show the coherence of our approximation. In that scenario, two resources have been added which results in the decrease of the collision probability. Therefore, with respect to the first scenario, Assumption 1 is more accurate and, the probability of being discontent decreases. Consequently, the numerical results of our approximation are closer to Monte Carlo simulations.

In addition, from figures \ref{fig:FEHT} and \ref{fig:FEHT_Cp2K}, one can check the result from proposition \ref{prop:TELprop}, in which the behaviour of \ac{EFHT} is $T_{EFHT}=\mathcal{O}(\frac{ 1}{\epsilon^{a_2}})$ where $a_2>0$ and the behaviour of the stability is $1-\alpha=\mathcal{O}(\epsilon^{a_4})$ where $a_4>0$.

%

\begin{figure}[t!]
    \centering
    \subfloat[\ac{ODL} : convergence time]{\includegraphics[width=0.4\textwidth]{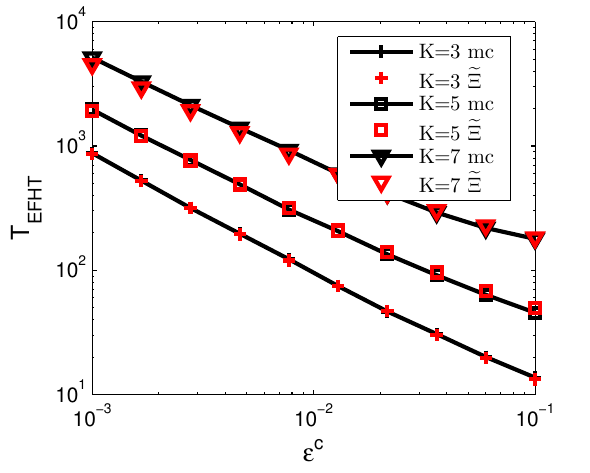}
        \label{fig:ODL_FEHT}}
    \hspace{0.5cm}
    \subfloat[\ac{ODL} : fraction of time]{	\includegraphics[width=0.4\textwidth]{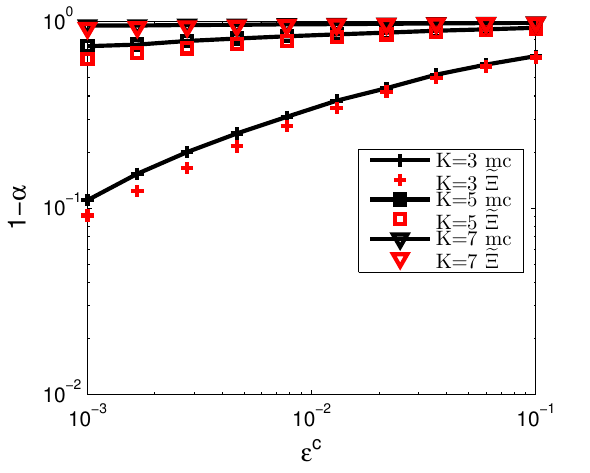}
        \label{fig:ODL_proptime}}
    \caption{\ac{EFHT} and stability comparison between our approximated models and Monte Carlo simulations when $N=K$.}
    \label{fig:ODL_CK}
        \vspace{-0.75cm}
\end{figure}

For \ac{ODL} algorithm, Figures \ref{fig:ODL_FEHT} and \ref{fig:ODL_proptime} present  when $K=N$ the \ac{EFHT} and the fraction of time $\alpha_j$, respectively. For both features, these results are accurate in comparison to Monte Carlo simulations. The gap observed at low $\epsilon$ for stability metric is due to the number of discontent players. We recall that the proposed approximation models accurately at most two discontent players. When $\epsilon$ decreases the number of discontent players increases (a player remains in D with probability $1-\epsilon$ when $u=0$) above two with an increasing probability and the model is less accurate. 

\begin{figure}[t!]
    \centering
    \subfloat[\ac{ODL} : convergence time]{\includegraphics[width=0.4\textwidth]{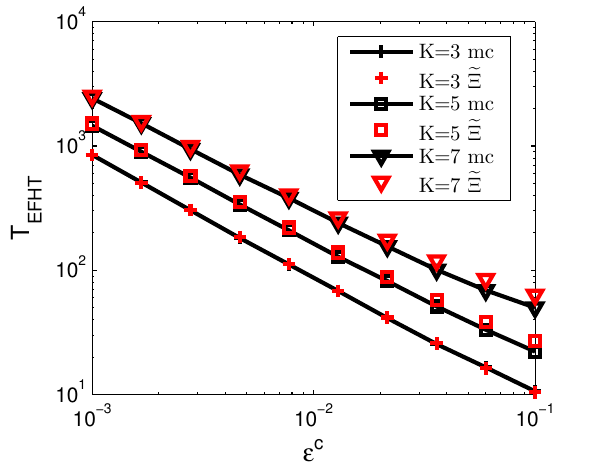}
        \label{fig:ODL_FEHT_Cp2K}}
    \hspace{0.5cm}
    \subfloat[\ac{ODL} : fraction of time]{	\includegraphics[width=0.4\textwidth]{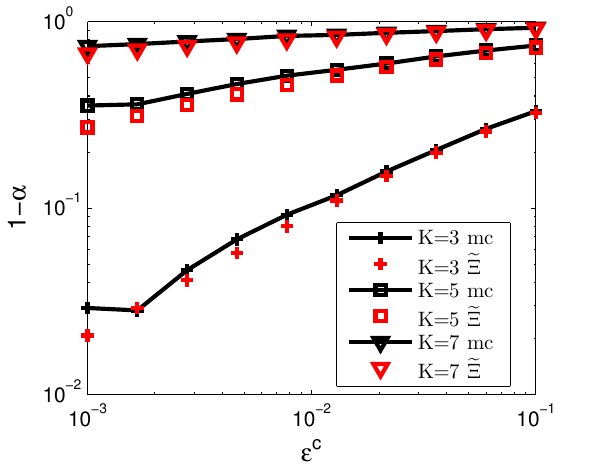}
        \label{fig:ODL_proptime_Cp2K}}
    \caption{\ac{EFHT} and stability comparison between our approximated models and Monte Carlo simulations when $N=K+2$.}
    \label{fig:ODL_Cp2K}
        \vspace{-0.75cm}
\end{figure}

We present in Fig. \ref{fig:ODL_FEHT_Cp2K} and~\ref{fig:ODL_proptime_Cp2K} the same results but with $N=K+2$. The accuracy of both features studied is again assessed. The probability to have collisions decreases and so does the probability to have an high number of discontent players. This leads to a better accuracy of the proposed model.

Generally, one can notice how the stability decreases with the number of players and how the convergence time increases. Furthermore, the convergence time decreases when the number of resource increases. In addition, one can check the results from 

In addition, from figures \ref{fig:ODL_CK} and \ref{fig:ODL_Cp2K}, one can check the result from proposition \ref{prop:ODLprop}, in which the behaviour of \ac{EFHT} is $T_{EFHT}=\mathcal{O}(\frac{ 1}{\epsilon^{c b_2}})$ where $b_2>0$ and the behaviour of the stability is $1-\alpha=\mathcal{O}(\epsilon^{c b_4})$ where $b_4>0$.

\subsection{Performance comparisons with approximation in the literature}
In this section, we compare the results obtained in previous section with the approximation given in paper~\cite{rose2014self}, noted model 1 in this work.  This last model figures of merit are computed as follows. For the \ac{EFHT}, the equation (33) in \cite{rose2014self} is employed. For the stability, Theorem 5 in \cite{rose2014self} gives the stability $\alpha_j$ but some corrections have been made. For instance, the term $T_{CNE}(k)$ from \cite{rose2014self} is replaced with equation (33)\cite{rose2014self} whose sum is started in $k$ instead of 0. The reason for this change is that the variable $T_{CNE}(k)$ diverges when $N=K$ and, it is an upper bound of (33) \cite{rose2014self}.

\begin{figure}[t!]
    \centering
    \subfloat[Convergence time]{\includegraphics[width=0.4\textwidth]{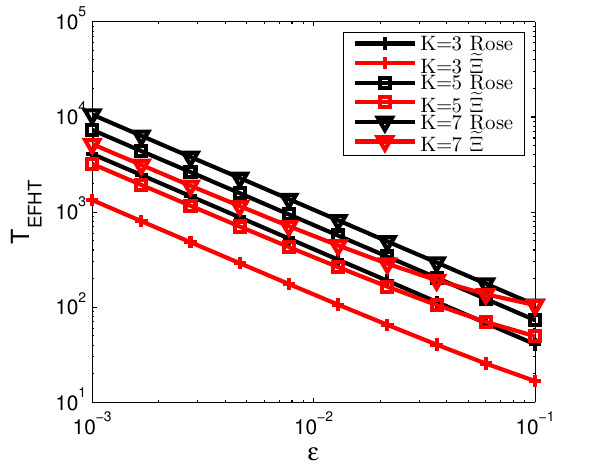}
        \label{fig:FETH_comparison_CK}}
    \hspace{0.5cm}
    \subfloat[Fraction of time]{\includegraphics[width=0.4\textwidth]{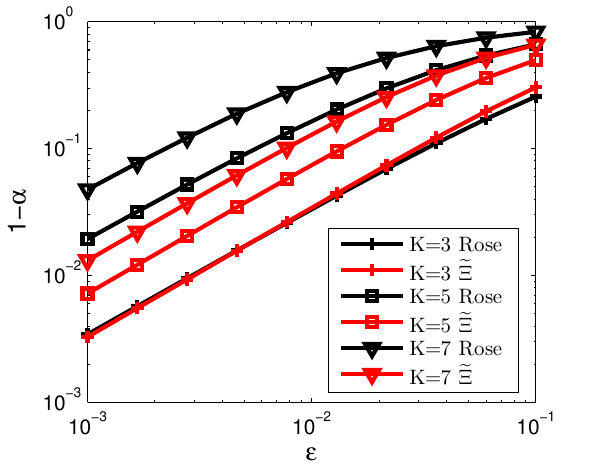}
        \label{fig:stab_comparison_CK}}
    \caption{Performance comparison between our approximation and Rose approximation when $N=K$.}
    \label{fig:comparison_CK}
        \vspace{-0.75cm}
\end{figure}
\begin{figure}[t!]
    \centering
    \subfloat[Convergence time]{\includegraphics[width=0.4\textwidth]{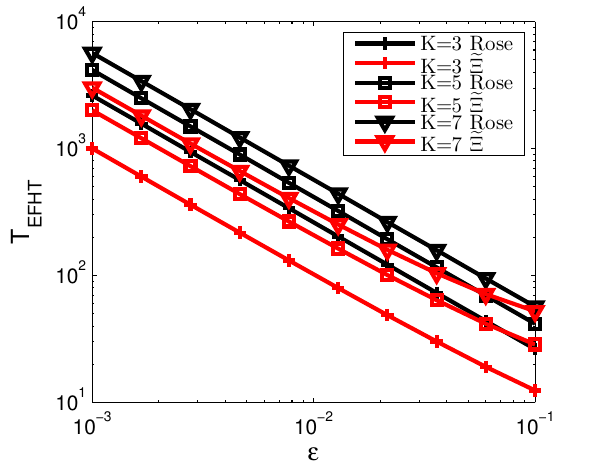}
        \label{fig:FETH_comparison_Cp2K}}
    \hspace{0.5cm}
    \subfloat[Fraction of time]{	\includegraphics[width=0.4\textwidth]{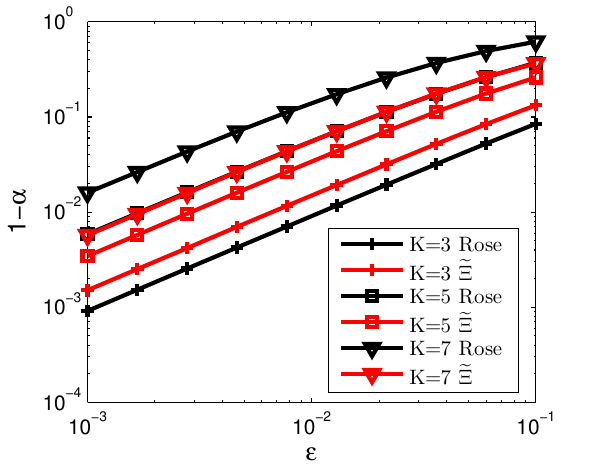}
        \label{fig:stab_comparison_Cp2K}}
    \caption{Performance comparison between our approximation and Rose approximation when $N=K+2$.}
    \label{fig:comparison_Cp2K}
    \vspace{-0.75cm}
\end{figure}
%
Figures \ref{fig:FETH_comparison_CK} and \ref{fig:FETH_comparison_Cp2K} present the \ac{EFHT} of the models $\widetilde{\Xi}$ and 1  when $N=K$ and $K+2$, respectively. One can observe that both models are quite far from each other except for high $\epsilon$. Knowing that our model converges close to simulations, we immediately deduce the model 1 lack of accuracy.


On the other hand, Figures \ref{fig:stab_comparison_CK} and \ref{fig:stab_comparison_Cp2K} present $1-\alpha_j$ when $N=K$ and $K+2$ respectively. One can notice that, except for $K=3$, the curves resulting from model 1 are above those of model $\widetilde{\Xi}$. As our model is a tight upper bound on Monte Carlo simulation (see figures \ref{fig:TE_proptime} and \ref{fig:TE_proptime_Cp2K} ), it again assesses the accuracy of our model.

 For the stability metric, in the case $N=3$ and $K=N$, the model 1 is as closed to Monte Carlo simulations as our proposed approximation. However, contrary to model $\widetilde{\Xi}$, the result for $N=K+2$ shows that model 1 gets away from the simulation contrary to our approximation that gets closer. This proves the coherence of our model in comparison to model 1.

\subsection{Performance comparison between \ac{TEL} and \ac{ODL}}

\begin{figure}[t!]
    \centering
    \subfloat[Convergence time]{\includegraphics[width=0.4\textwidth]{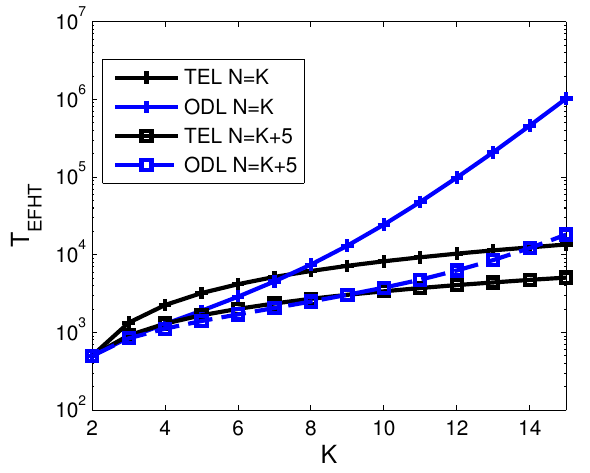}
        \label{fig:FETH_comparison_MTE_ODL}}\hspace{0.5cm}
    \subfloat[Fraction of time]{	\includegraphics[width=0.4\textwidth]{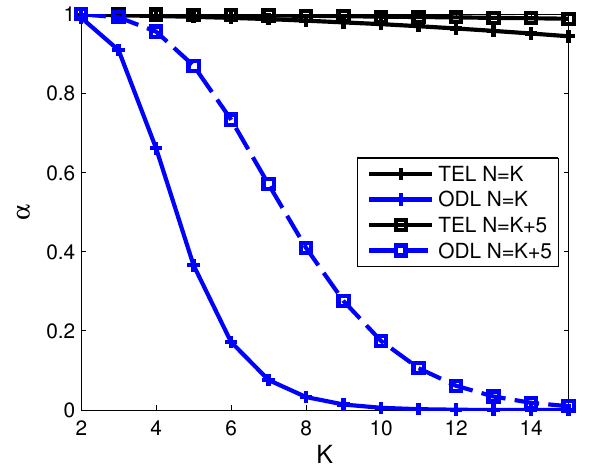}
        \label{fig:stab_comparison_MTE_ODL}}
    \caption{Performance comparison between \ac{TEL} and  \ac{ODL} when $\epsilon=10^{-3}$ with respect to $K$.}
    \label{fig:comparison_TEL_ODL}
        \vspace{-0.75cm}
\end{figure}

\begin{figure}[t!]
    \centering
    \subfloat[Convergence                         time]{\includegraphics[width=0.4\textwidth]{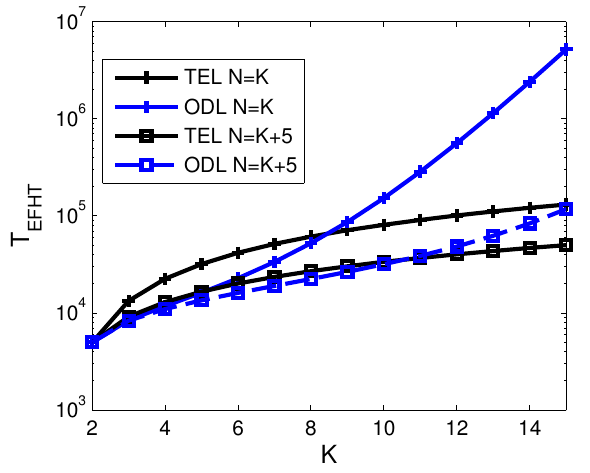}
        \label{fig:FETH_comparison_TEL_ODL_2}}\hspace{0.5cm}
    \subfloat[Fraction of time]{	\includegraphics[width=0.4\textwidth]{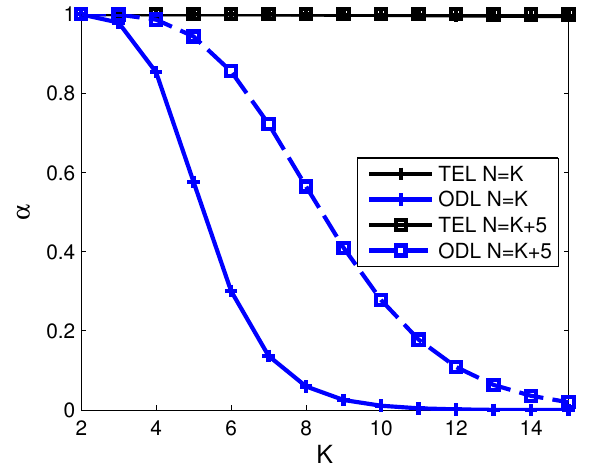}
        \label{fig:stab_comparison_TEL_ODL_2}}
    \caption{Performance comparison between \ac{TEL} and  \ac{ODL} when $\epsilon=10^{-4}$ with respect to $K$.}
    \label{fig:comparison_TEL_ODL_2}
        \vspace{-0.75cm}
\end{figure}

\begin{figure}[t!]
    \centering
    \subfloat[Fraction of time]{\includegraphics[width=0.4\textwidth]{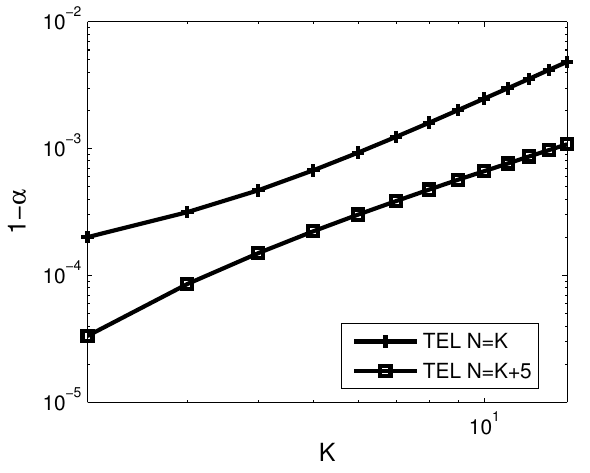}
        \label{fig:TEL_stabwithrespectK}}\hspace{0.5cm}
    \subfloat[Fraction of time]{	\includegraphics[width=0.4\textwidth]{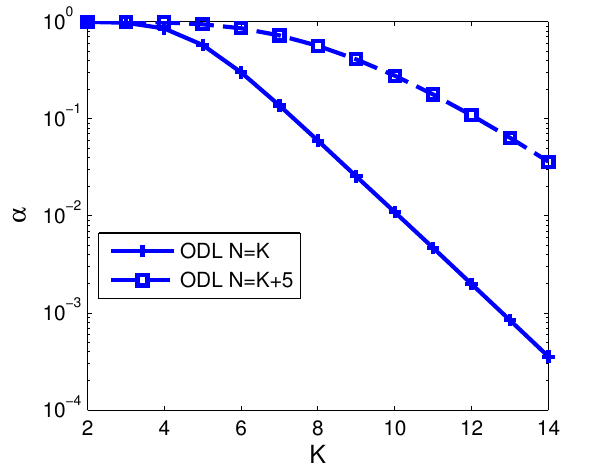}
        \label{fig:ODL_stab_withrespectK}}
    \caption{Stability performance of \ac{TEL} and  \ac{ODL} with respect to K.}
    \label{fig:TEL_ODL_stability}
    \vspace{-0.75cm}
\end{figure}

In previous sections, we have characterized the accuracy of \ac{TEL} and \ac{ODL} proposed models. In this section, we take the advantage of the available approximations that have low complexity, to compare both algorithms and, to analyse their performance in domains hardly reachable with Monte Carlo simulations. Figures \ref{fig:FETH_comparison_MTE_ODL} and \ref{fig:stab_comparison_MTE_ODL} present, for both algorithms, the \ac{EFHT} (in log scale) and the stability $\alpha_j$  respectively. The number of resources used is $N=K$ and $N=K+5$ and, the probability to experiment is fixed to $\epsilon=\epsilon^\nu=10^{-3}$. The  increase of $N$ results for both algorithms, first, in a better stability and, secondly, in a lower convergence time. This result counteracts the argument that the convergence time increases with the alphabet size (\cite{rose2014self} section V. B.).  The reason is that players find a free interference state faster and, the probability that two players collide is less important when the set of free resources is bigger. There exists a value of $K$ such that the \ac{EFHT} of both algorithms is the same. Below this value \ac{ODL} is more efficient than \ac{TEL} with respect to the convergence time and beyond this value the behavior is inverted. More generally, the fact that in some cases \ac{TEL} converges faster than \ac{ODL} contradicts the idea that, the bigger is the algorithm controller (4 moods for \ac{TEL} and 2 moods for \ac{ODL}), the slower its convergence is, as it is said in \cite{sheng2014utility} (section IV-B). In addition, Figure \ref{fig:stab_comparison_MTE_ODL} shows that \ac{TEL} is much more stable than \ac{ODL} even when $N$ is increased and such, at any $K$. Figure \ref{fig:comparison_TEL_ODL_2} presents the same results as in Figure \ref{fig:comparison_TEL_ODL} but with $\epsilon=\epsilon^\nu=10^{-4}$. The convergence time of both algorithms are increased. This is not a surprise as we deduce from proposition \ref{prop:TELprop} that $T_{EFHT}=\mathcal{O}\left(\frac{1}{\epsilon^{a_1}} \right)$. The decrease of $\epsilon$ increases the stability of both algorithms. As $\epsilon$ decreases, so does the number of experiments from players in state C. Thus, the probability that two players or more collide also decreases with $\epsilon$ which results in an higher stability of the state. More generally, the convergence and stability tendency remain the same in comparison to Figure \ref{fig:comparison_TEL_ODL}. From figures \ref{fig:FETH_comparison_MTE_ODL} and \ref{fig:FETH_comparison_TEL_ODL_2} one can assess what results from observations \ref{prop:TELprop} and \ref{prop:ODLprop}. The \ac{EFHT} of \ac{ODL} and \ac{TEL} respectively follow an exponential and a polynomial behaviour with respect to $K$ (the y-axis is in log scale). From figures \ref{fig:stab_comparison_MTE_ODL} and \ref{fig:stab_comparison_TEL_ODL_2} one can guess the exponential and polynomial decreasing of the \ac{ODL} and \ac{TEL} stability respectively. Figure \ref{fig:TEL_ODL_stability} presents the stability $1-\alpha$ and $\alpha$ with respect to $K$ for the \ac{TEL} and \ac{ODL} respectively. These two figures confirm the previous guess and assess the convergence results of observations \ref{prop:TELprop} and \ref{prop:ODLprop}. 

To conclude, in our system model, \ac{ODL} is less stable than \ac{TEL}. There exists some region of $K$ for which \ac{ODL} converges faster. However, the gain in speed convergence is not considerable and, the exponential behavior of \ac{ODL} with respect to K makes the convergence of this algorithm possibly very long in large systems. This small advantage in convergence speed is compromised by less stability. In view of the results, I would recommend that, the use of  \ac{ODL} algorithm in an environment with important utility variation is preferable when the need in stability is not important and the amount of players is limited.

%

\section{Conclusion}
\label{sec:conclusion}
This work provides a detailed performance analysis of well known  model-free learning strategies, \ac{TEL} and \ac{ODL}, that converge in a broad class of games.  To overcome the huge dimension of the inherent Markov Chains of the game, we provide an approximation of these chains. This allows computing a close approximation of the average time the system stays in a desired state as well as the average time required to achieved that state for the first time. Thanks to the above approximations, a comparison between the performance of \ac{TEL} and \ac{ODL} is provided.  

\appendices
\section{Probabilities involved in \ac{TEL} approximation}
\label{app:probTEL}

This appendix describes all the possible transitions and probabilities of \ac{TEL} algorithm model presented in Figures~\ref{fig:MC_TE}.

Remind with hypothesis \ref{hyp:hyp1} that, a discontent player accepts a free resource with probability $\epsilon^0=1$ and, it accepts a resource already interfered with probability $\epsilon^{F(0)}=\epsilon^{\frac{1}{2K}}$. In addition, with assumption \ref{ass:ass1}, we approximate the probability that there is one experimentation among $K$ content players by the probability that at least one experimentation happens $P_\epsilon(K)=1-(1-\epsilon)^K$.

\subsection{Notations and preliminaries}
\label{subsec:notations}

Like in Figure \ref{fig:MC_TE}, we consider two sets $\xi^n(i)$ and $\xi^{n+1}(j)$. We assume the presence of all the intermediary states in order to derive the most general transition probabilities. In practice, using section \ref{subsec:TELmodel}, the reader must check the existence of the intermediary states involved in sets before computing the probabilities. 

The probability computation require the knowledge of the players repartition over resources. For each set $\xi^n(i)$, the number of resources having $p$ players in $\bZ_n(i)$ is noted $M_{n}^i(p)=\sum_k \ind{S_{n,k}^i=p}$ and the number of players that share a resource with $p-1$ other players in $\bZ_n(i)$ is noted $m_{n}^i(p)=p M_{n}^i(p)$.

During a transition from $\xi^n(i)$ to $\xi^{n+1}(j)$, a resource is decremented by one player and a free resource is incremented. We note $k(i,j)$ the resource decremented such that the term $S_{n,k(i,j)}^i$ of $\bS_n(i)$ is decremented by one. Meanwhile, the term $S_{n,n+1}^i$ is incremented by one. During the reverse transition from $\xi^{n+1}(j)$ to $\xi^n(i)$, the most left column of $\bS_{n+1}(j)$ that has $S_{n,k(i,j)}^i-1$ players is incremented by one and, $S_{n+1,n+1}^j$ is decremented by one.

\subsection{Transitions inside each $\xi^n(i)$}
\label{subsubsec:insidexinTEL}
We start by describing the transitions inside $\xi^n(i)$ that is to say between the states  $\bZ_n(i)$, $\xi_0^n(i)$, $\xi_1^n(i)$, $\xi_2^n(i)$, $\xi_3^n(i)$ and $\xi_4^n(i)$. State $\bZ_n(i)$ is connected to $\xi_1^n(i)$ and itself. The transition $\bZ_n(i)\rightarrow \xi_1^n(i)$ happens when a player that is not interfered becomes watchful. It is given by probability
\begin{equation}
p_{\bZ_n(i) \xi_1^n(i)}=\overbrace{P_\epsilon(K)\frac{K-1}{K}}^{(a)}\overbrace{\frac{M_n^i(1)}{N-1}}^{(b)}.
\label{eq:Znitoxi1niTEL}
\end{equation}
where (a) is the probability that there is an experimentation from any player except the one that is going to be interfered and, (b) is the probability to select the frequency of a player not interfered. The probability of transition $\bZ_n(i)\rightarrow \bZ_n(i)$ is computed using the conservation probability property
\begin{equation}
p_{\bZ_n(i)\bZ_n(i)}=1-p_{\bZ_n(i) \xi_1^n(i)}-p_{\bZ_n(i)\bZ_{n+1}},
\label{eq:znitozniTEL}
\end{equation}
where $p_{\bZ_n(i)\bZ_{n+1}}$ is the probability for the network to find a new resource. It is given by
\begin{equation}
p_{\bZ_n(i) \bZ_{n+1}}=\overbrace{P_\epsilon(K) \frac{(K-m_n^i(1))}{K}}^{(a)}\overbrace{ \frac{M_i^n(0)}{N-1}}^{(b)},
\label{eq:total_prob1}
\end{equation}
where (a) is the probability that an interfered player experiments and, (b) is the probability that it finds a free resource.

From the state $\xi_1^n(i)$, the network can directly go in $\bZ_n(i)$ or $\xi_2^n(i)$. With assumption \ref{ass:ass1}, we do not consider any experimentation in $\xi_1^n(i)$ except the one needed to make the Markov chain ergodic in transition $\xi_1^n(i) \rightarrow \xi_2^n(i)$.  This later happens if the watchful player (which cannot experiment) is subjected to a second experiment on its resource by an other player. This is given by the following probability 
\begin{equation}
p_{\xi_1^n(i) \xi_2^n(i)}=\mathcal{P}_\epsilon(K-1)\frac{1}{N-1}.
\label{eq:xi1nitoxi2niTEL}
\end{equation}
Otherwise, we do not consider any other case from $\xi_1^n(i) $ and, the system goes naturally from $\xi_1^n(i) $ to $\bZ_n(i)$ with probability 

\begin{equation}
p_{\xi_1^n(i) \bZ_n(i)}=1- p_{\xi_1^n(i) \xi_2^n(i)}.
\label{eq:xi1nitozniTEL}
\end{equation}
State $\xi_2^n(i)$ is connected to $\bZ_n(i)$ and $\xi_3^n(i)$. During transition $\xi_2^n(i) \rightarrow \bZ_n(i)$, on the first hand, the discontent player selects  a free resource. There are $M_i^n(0)$ free resources in addition to the discontent player resource. Secondly, it accepts it as a new benchmark with probability $\epsilon^{F(1)}=1$ (hypothesis 2). The probability of this transition is thus given by 
\begin{equation}
p_{\xi_2^n(i) \bZ_n(i)}=\frac{M_i^n(0)+1}{N},
\label{eq:xi2nitoZniTEL}
\end{equation}
The transition $\xi_2^n(i) \rightarrow \xi_3^n(i)$ happens if the discontent player selects a resource already occupied by only one player and, it updates its benchmark with probability $\epsilon^{F(0)}=\epsilon^{\frac{1}{2K}}$ (hypothesis \ref{hyp:hyp1}). The new player interfered becomes watchful in next iteration. From the discontent player point of view, there are $M_i^n(1)-1$ players alone on their resource. The transition happens with the following probability 
\begin{equation}
p_{\xi_2^n(i) \xi_3^n(i)}=\frac{M_i^n(1)-1}{N}\epsilon^{\frac{1}{2K}} .
\label{eq:xi2nixi3niTEL}
\end{equation}
The probability of transition $\xi_2^n(i)\rightarrow \xi_2^n(i)$ is computed using the conservation probability property
\begin{equation}
p_{\xi_2^n(i)\xi_2^n(i)}=1-p_{\xi_2^n(i)\bZ_n(i)}-p_{\xi_2^n(i)\xi_3^n(i)}-p_{\xi_2^n(i)\bZ_{n-1}},
\label{eq:xi2nitoxi2niTEL}
\end{equation}
where $p_{\xi_2^n(i)\bZ_{n-1}}$ is the probability that the discontent player selects and accepts a resource already occupied by two players or more, which has probability
\begin{equation}
p_{\xi_2^n(i)\bZ_{n-1}}=\frac{N-M_n^i(1)-M_n^i(0)}{N} \epsilon^{\frac{1}{2K}}
\end{equation}

In state $\xi_3^n(i)$, the system is not aligned and there is no discontent player. Thus, with assumption~\ref{ass:ass1}, no experiment is proceeded and, the system moves directly to state $\xi_4^n(i)$ with probability
\begin{equation}
p_{\xi_3^n(i)\xi_4^n(i)}=1.
\label{eq:xi3nitoxi4niTEL}
\end{equation} 
The state $\xi_4^n(i)$ is connected to $\xi_0^n(i)$ and $\xi_3^n(i)$.  For the following transitions, it useful to note that in $\xi_4^n(i)$, the number of resources with players that do not interfere is $M_i^n(1)-2$ and, the number of free resources is $M_i^n(0)+1$. The transition $\xi_4^n(i) \rightarrow \xi_0^n(i)$ corresponds to the situation where the discontent player chooses a free resource, that it accepts with probability $\epsilon^{F(1)}=1$ (hypothesis~\ref{hyp:hyp1}). The player left alone sees its utility increases and becomes hopeful. This happens with probability 
\begin{equation}
p_{\xi_4^n(i) \xi_0^n(i)}=\frac{M_i^n(0)+1}{N}.
\label{eq:xi4nitoxi0niTEL}
\end{equation}
Transition $\xi_4^n(i)\rightarrow\xi_3^n(i)$ occurs when the discontent player selects an occupied resource with one player (there are $M_i^n(1)-2$ of them from $\xi_4^n(i)$) and, it accepts this resource as a new benchmark with probability $\epsilon^{\frac{1}{2 K}}$ according to hypothesis~\ref{hyp:hyp1}. Thus, the new player interfered becomes watchful with probability
\begin{equation}
p_{\xi_4^n(i) \xi_3^n(i)}=\frac{M_i^n(1)-2}{N} \epsilon^{\frac{1}{2K}}.
\label{eq:xi4nitoxi3niTEL}
\end{equation}
The probability to remain in $\xi_4^n(i)$ is 
\begin{equation}
p_{\xi_4^n(i) \xi_4^n(i)}=1-p_{\xi_4^n(i) \xi_0^n(i)}-p_{\xi_4^n(i) \xi_3^n(i)}-p_{\xi_4^n(i) \bZ_{n-1}},
\label{eq:xi4nitoxi4niTEL}
\end{equation}
where $p_{\xi_4^n(i) \bZ_{n-1}}$ is the probability that, the network use one less frequency with all players content and aligned. This happens if the discontent player selects and accepts one of the $N-M_n^i(1)-M_n^i(0)$ resources already occupied by two players or more or, if it selects and accepts its current resource where the content player interfered is aligned (i.e. it is going to accept the choice of the discontent player). Consequently, the system uses one less frequency with probability 
\begin{equation}
p_{\xi_4^n(i) \bZ_{n-1}}=\frac{N-M_n^i(1)-M_n^i(0)+1}{N} \epsilon^{\frac{1}{2K}},
\end{equation}
and the probability that it remains one more step in $\xi_4^n(i)$ is $p_{\xi_4^n(i) \xi_4^n(i)}=0$.
Once the network is in $\xi_0^n(i)$, one player is hopeful and with assumption 1, no player can experiment. Thus, in next step this player becomes content with a benchmark update and the network goes to $\bZ_n(i)$ with the following probability

\begin{equation}
p_{\xi_0^n(i)\bZ_n(i)}=1.
\label{eq:xi0nitozniTEL}
\end{equation}

\subsection{Transitions from $\xi^n(i)$ to $\xi^{n+1}(j)$}
The only way for the network to find a new resource is to go through a \ac{RC} $\bZ_n(i)$. If the network is not in $\bZ_n(i)$ either one player is discontent or the network is not aligned. In the later, assumption 1 tells us that no player experiments, whereas in the former, the discontent player cannot discover a free resource because it is not interfered in $\bZ_n(i)$.

During transition $\bZ_n(i)\rightarrow \bZ_{n+1}(j)$, a player experiments on a free resource with probability
\begin{equation}
p_{\bZ_n(i) \bZ_{n+1}(j)}=
\begin{cases}
& \overbrace{P_\epsilon(K) \frac{m_n^i(S_{n,k(i,j)}^i)}{K}}^{(a)}  \overbrace{\frac{M_i^n(0)}{N-1}}^{(b)}, \text{ if }S_{n,k(i,j)}^i>2,\\
& 0, \text{ if }S_{n,k(i,j)}^i=2, 
\end{cases}
\label{eq:znitoznp1jTEL}
\end{equation}
where (a) is the probability to have an experimentation from any player interfered on resources with $S_{n,k(i,j)}^i>2$ players and, (b) is the probability to select a free resource. The second line corresponds to an other transition $\bZ_n(i)\rightarrow \xi_0^{n+1}(j)$, in which the player left alone after the experimentation sees its utility increase and becomes hopeful. The probability of transition $\bZ_n(i)\rightarrow \xi_0^{n+1}(j)$ is thus complementary to the previous one and, it is given by
\begin{equation}
p_{\bZ_n(i) \xi_0^{n+1}(j)}=
\begin{cases}
& 0, \text{ if }S_{n,k(i,j)}^i>2,\\
& P_\epsilon(K) \frac{m_n^i(S_{n,k(i,j)}^i)}{K} \frac{M_i^n(0)}{N-1}, \text{ if }S_{n,k(i,j)}^i=2.
\end{cases}
\label{eq:znitoxi0np1jTEL}
\end{equation}
\subsection{Transitions from $\xi^{n+1}(j)$ to $\xi^{n}(i)$} 

The approximation is constructed such that if it is possible to go from $\xi^{n}(i)$ to $\xi^{n+1}(j)$, it is also possible to go from $\xi^{n+1}(j)$ to $\xi^{n}(i)$ (see section \ref{subsec:TELmodel}). The way for the system to go in a set where one less resource is employed only happens in states with a discontent player, i.e.  $\xi_2^{n+1}(j)$ and $\xi_4^{n+1}(j)$ for  transition $\xi^{n+1}(j)$ to $\xi^{n}(i)$. In practice, to compute the transitions inside $\xi^{n+1}(j)$, we use the formulas in appendix \ref{subsubsec:insidexinTEL} by replacing the indices appropriately. In this section, the starting state is  in  $\xi^{n+1}(j)$. Therefore, we use the functions $m_{n+1}^j(.)$ and $M_{n+1}^j(.)$ instead of $m_{n}^i(.)$ and  $M_{n}^i(.)$. Moreover, during a transition from  $\xi^{n}(i)$ to $\xi^{n+1}(j)$, the resource that contained $S_{n,k(i,j)}^{i}$ in $\bS_{n}(i)$ has been decremented by one. Thus, the transition from $\xi^{n+1}(j)$ to $\xi^{n}(i)$ occurs if any resource that contains $S_{n,k(i,j)}^{i}-1$ players is incremented by one.

The transition $\xi_2^{n+1}(j) \rightarrow \bZ_{n}(i)$ happens if the discontent player selects a frequency with $S_{n,k(i,j)}^{i}-1$ players and accept the new benchmark. The probability of $\xi_2^{n+1}(j) \rightarrow \bZ_{n}(i)$ is thus given by
\begin{equation}
p_{\xi_2^{n+1}(j)  \bZ_{n}(i)}=
\begin{cases}
& \frac{M_j^{n+1}(S_{n,k(i,j)}^{i}-1)}{N}\epsilon^{\frac{1}{2K}},\text{ if } S_{n,k(i,j)}^{i}-1\geq2,\\
& 0,\text{ if } S_{n,k(i,j)}^{i}-1=1,\\
\end{cases}
\label{eq:xi2np1jtozniTEL}
\end{equation}
where, the second line is null because it is represented by transition  $\xi_2^{n+1}(j) \rightarrow \xi_3^{n+1}(j)$ (see \eqref{eq:xi2nixi3niTEL} with appropriate indices changes).

The transition $\xi_4^{n+1}(j)\rightarrow\bZ_{n}(i)$ happens when the discontent player selects an occupied resource and all players are content and aligned in the end. This is given by probability
\begin{equation}
p_{\xi_4^{n+1}(j) \bZ_{n}(i)}=
\begin{cases}
& 0,\text{ if } S_{n,k(i,j)}^{i}-1\geq 2,\\
& \frac{1}{N}\epsilon^{\frac{1}{2K}},\text{ if } S_{n,k(i,j)}^{i}-1=1,
\end{cases}
\label{eq:xi4np1jtozniTEL}
\end{equation}
where, the first line is null because this corresponds to the transition $\xi_4^{n+1}(j)\rightarrow \xi_0^{n}(i)$ described afterwards. The second line is the probability for the discontent player to select and to accept the current resource. The transition  $\xi_4^{n+1}(j)\rightarrow \xi_0^{n}(i)$ corresponds to the case where the discontent player selects and accepts a resource with $S_{n,k(i,j)}^{i}-1 \geq 2$ players. Consequently, the player that is left alone becomes hopeful with probability
\begin{equation}
p_{\xi_4^{n+1}(j) \xi_0^{n}(i)}=
\begin{cases}
&\frac{M_j^{n+1}(S_{n,k(i,j)}^{i}-1)}{N}\epsilon^{\frac{1}{2K}},\text{ if } S_{n,k(i,j)}^{i}-1\geq 2,\\
& 0,\text{ if } S_{n,k(i,j)}^{i}-1=1,\\
\end{cases}
\label{eq:xi4np1jtoxi0niTEL}
\end{equation}
where, the second line corresponds to previous transition $\xi_4^{n+1}(j)\rightarrow\bZ_{n}(i)$.

\section{Probabilities involved in \ac{ODL} approximation}
\label{app:probODL}
This appendix describes all the possible transitions and probabilities of \ac{ODL} algorithm model presented in Figure~\ref{fig:MCmodODL}. We also use the same notations and preliminaries specified in appendix \ref{subsec:notations}.

In \ac{ODL}, a player which perceives a utility or an action change accepts the new benchmark with probability $\epsilon^{1-u}$ or, it refuses it and becomes discontent with probability $1-\epsilon^{1-u}$.

\subsection{Transitions inside $\xi^n(i)$} 
\label{subsubsec:transitionZnxinODL}
We  start by describing the transitions between the states $\bZ_n(i)$, $\xi_1^n(i)$, $\xi_2^n(i)$ and $\xi_3^n(i)$. The state  $\bZ_n(i)$ is connected to $\xi_1^n(i)$, $\xi_2^n(i)$, $\xi_3^n(i)$.  Transition $\bZ_{n}(i) \rightarrow \xi_1^n(i)$ happens if an alone player becomes discontent after perceiving a utility change that it does not accept. This situation arises with probability 

\begin{equation}
\begin{aligned}
p_{\bZ_{n}(i) \xi_1^n(i)}&=\overbrace{\mathcal{P}_\epsilon(K)\frac{(K-m_n^i(1)-m_n^i(2))}{K}\frac{M_n^i(1)}{N-1}}^{(a)}  \overbrace{(1-\epsilon)}^{(b)}.\\ 
\end{aligned}
\label{eq:znitoxi1inODL}
\end{equation}
where (a) is the probability that any player interfered by two players or more experiments on a resource with solely one player, (b) is the probability that this alone player becomes discontent.

The transition $\bZ_n(i) \rightarrow \xi_2^n(i)$ represents the situation where two players alone in $\bZ_n(i)$ become discontent in one step, whose probability is 
\begin{equation}
p_{\bZ_{n}(i) \xi_2^n(i)}=\overbrace{\mathcal{P}_\epsilon(K) \frac{m_n^i(1)}{K} \frac{M_n^i(1)-1}{N-1}}^{(a)} \overbrace{(1-\epsilon)^2}^{(b)},
\label{eq:znitoxi2niODL}
\end{equation}

where, (a) is the probability that an alone player experiments on a resource with an other alone player and, (b) is the probability that both players become discontent.

The transition $\bZ_n(i) \rightarrow \xi_3^n(i)$ represents the situation where, from a resource with two players, one of them experiments on an other resource with one player and, one of them ends in discontent mood. This happens with probability
\begin{equation}
p_{\bZ_{n}(i) \xi_3^n(i)}=\overbrace{\mathcal{P}_\epsilon(K) \frac{m_n^i(2)}{K}}^{(a)} \overbrace{\frac{M_n^i(1)}{N-1} 2 \epsilon (1-\epsilon)}^{(b)},
\label{eq:znitoxi3niODL}
\end{equation}

where (a) is the probability that a player experiments from a resource with two of them, (b) is the probability to interfere with one player and, one of the two players involved becomes discontent. The presence of multiplier 2 in term (b) means that, inverting player's label is a different event that 
results in the same state $\xi_3^n(i)$ and with the same probability.

The transition $\bZ_n(i) \rightarrow \bZ_n(i)$ is computed using probability conservation as follows,
\begin{equation}
\begin{aligned}
p_{\bZ_{n}(i) \bZ_{n}(i)}&=1-p_{\bZ_{n}(i) \xi_1^n(i)}-p_{\bZ_{n}(i) \xi_2^n(i)}-p_{\bZ_{n}(i)\xi_3^n(i)}-p_{\bZ_{n}(i) \bZ_{n+1}}-p_{\bZ_{n}(i) \bZ_{n-1}}\\
&-p_{\bZ_{n}(i)\xi^{n+1}_1}-p_{\bZ_{n}(i)\xi^{n+1}_2}-p_{\bZ_{n}(i)\xi^{n-1}_3},
\end{aligned}
\label{eq:znitozniODL}
\end{equation}
where $p_{\bZ_{n}(i) \bZ_{n+1}}$ is the probability for any interfered player to select a free resource, $p_{\bZ_{n}(i) \bZ_{n-1}}$ is the probability for any not interfered player to become interfered, $p_{\bZ_{n}(i)\xi^{n+1}_1}$ and $p_{\bZ_{n}(i)\xi^{n+1}_2}$ are similar to $p_{\bZ_{n}(i) \bZ_{n+1}}$ with one and two players ending in discontent mood respectively and, $p_{\bZ_{n}(i)\xi^{n-1}_3}$ is the probability that two players alone in $\bZ_n(i)$ finish on the same resource with one of them discontent. The first probability is given by
\begin{equation}
p_{\bZ_n(i) \bZ_{n+1}}=P_\epsilon(K) \frac{(K-m_n^i(1))}{K} \frac{M_i^n(0)}{N-1},
\label{eq:ZntoZnp1ODL}
\end{equation}
which is the same as equation~\eqref{eq:total_prob1} in the \ac{TEL} model. The second probability is given by
\begin{equation}
p_{\bZ_{n}(i) \bZ_{n-1}}=\overbrace{\mathcal{P}_\epsilon(K) \frac{m_n^i(1)}{K}}^{(a) }\bigg( \overbrace{\frac{N-M_n^i(1)-M_n^i(0)}{N-1}\epsilon}^{(b)}+ \overbrace{\frac{M_n^i(1)-1}{N-1} \epsilon^2}^{(c)} \bigg),
\label{eq:probaZntoZnm1ODL}
\end{equation}
where, (a) is the probability for a player that is not interfered to experiment, (b) is the probability that it experiments on resource with two players or more and that it updates its benchmark, (c) is the probability to select the resource of a player not interfered and that both accept this new benchmark.
Probability $p_{\bZ_{n}(i)\xi^{n+1}_1}$ is given by
\begin{equation}
p_{\bZ_n(i) \xi^{n+1}_1}=P_\epsilon(K) \frac{K-m_n^i(1)}{K} \overbrace{\frac{N-M_n^i(1)-M_n^i(0)-1}{N-1}(1-\epsilon)}^{(a)},
\label{eq:Zntoxi1np1ODL}
\end{equation}
where (a) is the probability that the player interfered selects a frequency with two players or more, except its own resource, and, that it ends in discontent mood.

The probability to end in $\xi^{n+1}_2$ is 
\begin{equation}
p_{\bZ_n(i) \xi^{n+1}_2}=P_\epsilon(K) \frac{K-m_n^i(1)}{K} \overbrace{\frac{M_n^i(1)}{N-1}(1-\epsilon)^2}^{(a)},
\label{eq:Zntoxi2np1ODL}
\end{equation}
where (a) is the probability that the experimenter selects the resource of player not interfered and that both end up in discontent mood.

Finally, the probability to go from $\bZ_n(i)$ to $\xi^{n-1}_3$ is given by
\begin{equation}
p_{\bZ_n(i) \xi^{n-1}_3}=P_\epsilon(K) \frac{m_n^i(1)}{K} \overbrace{\frac{M_n^i(1)-1}{N-1}2\epsilon(1-\epsilon)}^{(a)},
\label{eq:Zntoxi3nm1ODL}
\end{equation}
where (a) is the probability that the experimenter selects a resource with a player not interfered and, one of them ends in discontent mood. The multiplier 2 has a similar role than in \eqref{eq:znitoxi3niODL}.

The state $\xi_1^n(i)$ is connected to $\bZ_n(i)$ and $\xi_2^n(i)$ inside the set $\xi^n(i)$. During transition $\xi_1^n(i) \rightarrow \bZ_{n}(i)$ the discontent player either chooses a free resource or its current benchmark with probability 
\begin{equation}
p_{\xi_1^n(i) \bZ_{n}(i)}=\frac{M_n^i(0)+1}{N}.
\label{eq:xi1nitozniODL}
\end{equation}

During transition $\xi_1^n(i) \rightarrow \xi_2^n(i)$, the discontent player makes an other player discontent  in addition to itself. This is given by probability
\begin{equation}
p_{\xi_1^n(i) \xi_2^n(i)}=\overbrace{\frac{M_n^i(1)-1}{N}}^{(a)} (1-\epsilon)^2,
\label{eq:xi1nitoxi2niODL}
\end{equation}
where (a) is the probability  that the discontent player selects a resource that contains a player alone, except its own resource. 

The probability $p_{\xi_1^n(i)\xi_1^n(i)}$ is obtained using the conservation property:
\begin{equation}
p_{\xi_1^n(i)\xi_1^n(i)}=1-p_{\xi_1^n(i)\bZ_n(i)}-p_{\xi_1^n(i)\xi_2^n(i)}-p_{\xi_1^n(i)\bZ_{n-1}}-p_{\xi_1^n(i)\xi_3^{n-1}},
\label{eq:xi1nitoxi1niODL}
\end{equation}
where $p_{\xi_1^n(i)\bZ_{n-1}}$ and $p_{\xi_1^n(i)\xi_3^{n-1}}$ are the probability for the system starting in $\xi_1^n(i)$ to end for all $j\in I_N(n-1)$ in states $\bZ_{n-1}(j)$ and $\xi_3^{n-1}(j)$ respectively. The first probability is given by
\begin{equation}
p_{\xi_1^n(i) \bZ_{n-1}}=\frac{N-M_n^i(0)-M_n^i(1)}{N} \epsilon+\frac{ M_n^i(1)-1}{N}\epsilon^2,
\label{eq:ProbaXi1toZnm1ODL}
\end{equation} 
which are similar to terms (b)+(c) in \eqref{eq:probaZntoZnm1ODL} except the choice is made over all resources as the player is in state D.


The second probability is given by
\begin{equation}
p_{\xi_1^n(i) \xi_3^{n-1}}=\frac{M_n^i(1)-1}{N}2 \epsilon (1-\epsilon),
\label{eq:Probaxi1toxi3nm1m1ODL}
\end{equation} 
which is similar to (a) in \eqref{eq:Zntoxi3nm1ODL} except that the choice is made among $N$ resources.

The state $\xi_2^n(i)$ is connected to $\xi_1^n(i)$ and $\bZ_n(i)$. The probability of transition $\xi_2^n(i)\rightarrow \bZ_n(i)$ is given by
\begin{equation}
p_{\xi_2^n(i)\bZ_n(i)}= \overbrace{\frac{M_n^i(0)+2}{N}}^{(a)} \overbrace{\frac{M_n^i(0)+1}{N}}^{(b)},
\label{eq:xi2nitozniODL}
\end{equation}
where (a) is the probability that the first discontent player selects a free resource. The number of free resource is $M_n^i(0)$ in addition to the 2 resources left by the discontent players. Term (b) is the probability that the other discontent player selects a free resource given that, the first discontent player has already selected a free resource.

The probability of a transition $\xi_2^n(i)\rightarrow \xi_1^n(i)$ is given by
\begin{equation}
p_{\xi_2^n(i)\xi_1^n(i)}=2 \frac{M_n^i(0)+2}{N} \overbrace{\frac{N-M_n^i(1)-M_n^i(0)}{N}(1-\epsilon)}^{(a)},
\label{eq:xi2nitoxi1niODL}
\end{equation}
where (a) is similar to the term (a) in \eqref{eq:Zntoxi1np1ODL} except there is one more resource available.

The probability to remain in $\xi_2^n(i)$ is given by probability conservation
\begin{equation}
p_{\xi_2^n(i)\xi_2^n(i)}=1-p_{\xi_2^n(i)\bZ_n(i)}-p_{\xi_2^n(i)\xi_1^n(i)}-p_{\xi_2^n(i)\bZ_{n-1}}-p_{\xi_2^n(i)\xi_1^{n-1}}-p_{\xi_2^n(i)\xi_2^{n-1}}-p_{\xi_2^n(i)\xi_3^{n-1}},
\label{eq:xi2nitoxi2niODL}
\end{equation}
where $p_{\xi_2^n(i)\bZ_{n-1}}$, $p_{\xi_2^n(i)\xi_1^{n-1}}$ and $p_{\xi_2^n(i)\xi_2^{n-1}}$ represent the probability that the system uses one less resource and, that, respectively, all player are content and aligned, one player ends discontent and two players end discontent. The probability $p_{\xi_2^n(i)\xi_3^{n-1}}$ corresponds to the event where two players not interfered end on the same resource with one of them discontent. A transition $\xi_2^n(i) \rightarrow \bZ_{n-1}$ happens if one of the two discontent players selects a resource already occupied and the system ends in an all content and aligned state. The probability of the first events is given by
\begin{equation}
p_{\xi_2^n(i)\bZ_{n-1}}=  \frac{M_n^i(0)+2}{N}\frac{M_n^i(1)-1}{N}\epsilon^2+2 \frac{M_n^i(0)+2}{N}\frac{N-M_n^i(1)-M_n^i(0)}{N} \epsilon,
\label{eq:xi2toZnm1ODL}
\end{equation}



The transition $\xi_2^n(i)\rightarrow\xi_1^{n-1}$ happens if one of the two discontent players selects a resource already occupied and the system ends with one player discontent. The probability of all these possible events is given by
\begin{equation}
\begin{aligned}
p_{\xi_2^n(i)\xi_1^{n-1}}&=2 \frac{N-M_n^i(1)-M_n^i(0)}{N} \epsilon \frac{N-M_n^i(1)-M_n^i(0)}{N} (1-\epsilon)\\
&+2 \frac{M_n^i(1)-2}{N} \epsilon^2 \frac{N-M_n^i(1)-M_n^i(0)+1}{N} (1-\epsilon),
\end{aligned}
\label{eq:xi2np1toxi1nODL}
\end{equation}
where the first terms in the sum deals with the cases in which one of the two discontent players accepts a resource with two players or more and, the second term concerns the case in which one of the two players selects a resource with one player solely.

The probability of transition from $\xi_2^{n}(i)$ to $\xi_2^{n-1}$ is the probability that one player updates its benchmark with a resource already occupied and that, the systems ends with two discontent players. It is given by
\begin{equation}
p_{\xi_2^{n}(i)\xi_2^{n-1}}= 2\frac{N-M_n^i(1)-M_n^i(0)}{N}\epsilon \frac{M_n^i(1)-2}{N}(1-\epsilon)^2+\frac{M_n^i(1)-2}{N} \epsilon^2  \frac{M_n^i(1)-3}{N} (1-\epsilon)^2,
\label{eq:xi2ntoxi2nm1ODL}
\end{equation}
where the first term deals with the case in which, one of the discontent players updates its benchmark with a resource that contains two players or more and, the second term concerns the scenario where both discontent players select a resource with one player. More specifically, in the second term, once the first discontent player has selected a resource with one player and, both have accepted the new benchmark, there is now one less resource with one player, i.e. $M_n^i(1)-3$.


The transition  $\xi_2^{n}(i)$ to $\xi_3^{n-1}$ happens if one of the discontent players finds a free resource and, the other selects the resource of a player not interfered. In this last situation one of the two players interfered becomes discontent. This is given by the following probability
\begin{equation}
p_{\xi_2^n(i) \xi_3^{n-1}}=\frac{M_n^i(0)+2}{N}\frac{M_n^i(1)-1}{N}2 \epsilon (1-\epsilon),
\label{eq:xi2ntoxi3nm1ODL}
\end{equation}

The state $\xi_3^n(i)$ is linked to $\bZ_n(i)$ and itself in the set $\xi_n(i)$. During transition $\xi_3^n(i) \rightarrow \bZ_n(i)$ the system comes back to the all content and aligned  state with probability
\begin{equation}
p_{\xi_3^n(i)\bZ_n(i)}=\overbrace{\frac{\epsilon}{N}}^{(a)}+\overbrace{\frac{M_n^i(1)}{N}\epsilon^2}^{(b)},
\label{eq:xi3nitozniODL}
\end{equation}
where (a) is the probability that the discontent player tries the current resource and that it updates its benchmark. Note that, the player interfered is already aligned from $\xi_3^n(i)$. Term (b) is the probability that the discontent player tries an other  resource with one player and both accept the new benchmark.

The probability to remain in $\xi_3^n(i)$ is given by 
\begin{equation}
p_{\xi_3^n(i)\xi_3^n(i)}=1-p_{\xi_3^n(i)\bZ_n(i)}-p_{\xi_3^n(i)\bZ_{n+1}}-p_{\xi_3^n(i)\xi_1^{n+1}}-p_{\xi_3^n(i)\xi_2^{n+1}},
\label{eq:xi3nitoxi3niODL}
\end{equation}
where $p_{\xi_3^n(i)\bZ_{n+1}}$, $p_{\xi_3^n(i)\xi_1^{n+1}}$ and $p_{\xi_3^n(i)\xi_2^{n+1}}$ are the  probabilities for the system to use one more frequency from $\xi_3^n(i)$ and, respectively, the system ends with all players content, one player discontent and two players discontent. Note that, these transitions lead to one unique state $j$ in $I_N(n+1)$, $\bZ_{n+1}(j)$, $\xi_1^{n+1}(j)$ and $\xi_2^{n+1}(j)$. The transition $\xi_3^n(i) \rightarrow \bZ_{n+1}(j)$ occurs if the system ends in an all content mood and aligned with one more frequency used after the experimentation. It happens if the discontent player chooses a free resource with probability

\begin{equation}
p_{\xi_3^n(i)\bZ_{n+1}(j)}=
\begin{cases}
& \frac{M_n^i(0)}{N}, \text{ if } S_{n,k(i,j)}=2,\\
& 0, \text{ otherwise}.
\end{cases}
\label{eq:xi3toZnp1ODL}
\end{equation}
The term $p_{\xi_3^n(i)\bZ_{n+1}}$ is the sum over all possible $j$ of $p_{\xi_3^n(i)\bZ_{n+1}(j)}$. Consequently, $p_{\xi_3^n(i)\bZ_{n+1}}= \frac{M_n^i(0)}{N}$.

The transition $\xi_3^n(i) \rightarrow \xi_1^{n+1}(j)$ occurs if the discontent cluster remains discontent. It happens with probability
\begin{equation}
p_{\xi_3^n(i)\xi_1^{n+1}(j)}=
\begin{cases}
& \frac{N-M_n^i(1)-M_n^i(0)-1}{N}(1-\epsilon), \text{ if } S_{n,k(i,j)}=2, \\
& 0, \text{ otherwise}.
\end{cases}
\label{eq:xi3toxi1np1ODL}
\end{equation}
where the first line is the probability that the discontent player experiments on a resource, with two players or more, except the current one and, that it remains discontent. After this event, the player left alone by the discontent player is no more interfered and accepts the new  benchmark with probability 1. The total probability to go in $\xi_1^{n+1}$ is $p_{\xi_3^n(i)\xi_1^{n+1}}=\frac{N-M_n^i(1)-M_n^i(0)-1}{N}(1-\epsilon)$.

The event that leads to transition $\xi_3^n(i) \rightarrow \xi_2^{n+1}(j)$ is realized if the discontent player experiments on an other resource with a cluster not interfered and both end in discontent. This happens with probability
\begin{equation}
p_{\xi_3^n(i)\xi_2^{n+1}(j)}=
\begin{cases}
& \frac{M_n^i(1)}{N} (1-\epsilon)^2,  \text{ if } S_{n,k(i,j)}=2,\\
&  0, \text{ otherwise}.
\end{cases}
\label{eq:xi3toxi2np1ODL}
\end{equation}
Using the same reasoning, $p_{\xi_3^n(i)\xi_2^{n+1}}=\frac{M_n^i(1)}{N} (1-\epsilon)^2$.

\subsection{Transition from $\xi^n(i)$ to $\xi^{n+1}(j)$}
The only states in $\xi^n(i)$ from which the system can use one more frequency are $\bZ_n(i)$ and $\xi_3^n(i)$. In other states, the discontent players are alone on their resource, which mean that they cannot discover a new one.

The transitions $\xi_3^n(i)\rightarrow \bZ_{n+1}(j)$, $\xi_3^n(i)\rightarrow \xi_1^{n+1}(j)$ and $\xi_3^n(i)\rightarrow \xi_2^{n+1}(j)$ have been described in equations \eqref{eq:xi3toZnp1ODL}, \eqref{eq:xi3toxi1np1ODL} and \eqref{eq:xi3toxi2np1ODL}. 

The transition $\bZ_n(i)\rightarrow \bZ_{n+1}(j)$ happens if a player on a resource with $S_{n,k(i,j)}^i$ experiments on a free resource. It is given by probability
\begin{equation}
p_{\bZ_n(i) \bZ_{n+1}(j)}=P_\epsilon(K) \frac{m_n^i(S_{n,k(i,j)}^i)}{K} \frac{M_i^n(0)}{N-1},
\label{eq:ZnitoZnp1jODL}
\end{equation}
which is term $j$ of the sum that gives the total probability $p_{\bZ_n(i)\bZ_{n+1}}$ \eqref{eq:ZntoZnp1ODL}. The term $K-m_n^i(1)$ in \eqref{eq:ZntoZnp1ODL} is decomposed as follows $\sum_{j}m_n^i(S_{n,k(i,j)}^i)=K-m_n^i(1)$.

The transition from $\bZ_n(i)\rightarrow \xi_1^{n+1}(j)$ corresponds to the term $j$ of the sum that gives probability $\bZ_n(i)\rightarrow \xi_1^{n+1}$ in \eqref{eq:Zntoxi1np1ODL}. Using the same procedure
\begin{equation}
p_{\bZ_n(i) \xi^{n+1}_1(j)}=P_\epsilon(K) \frac{m_n^i(S^i_{n,k(i,j)})}{K} \frac{N-M_n^i(1)-M_n^i(0)-1}{N-1}(1-\epsilon),
\label{eq:Znitoxi1np1jODL}
\end{equation}
Again with the same decomposition, probability of transition  $\bZ_n(i)\rightarrow \xi_2^{n+1}(j)$ is obtained using \eqref{eq:Zntoxi2np1ODL} as follows
\begin{equation}
p_{\bZ_n(i) \xi^{n+1}_2(j)}=P_\epsilon(K) \frac{m_n^i(S^i_{n,k(i,j)})}{K} \frac{M_n^i(1)}{N-1}(1-\epsilon)^2,
\label{eq:Znitoxi2np1jODL}
\end{equation}

\subsection{Transitions from $\xi^{n+1}(j)$ to $\xi^{n}(i)$} 
The system can employ one less resource when an alone player selects a resource already occupied as a new benchmark. 
These transitions are possible from states $\bZ_{n+1}(j)$, $\xi_1^{n+1}(j)$ and $\xi_2^{n+1}(j)$. In practice, probability transitions inside $\xi^{n+1}(j)$, are computed from formulas in section \ref{subsubsec:transitionZnxinODL} by replacing the indices appropriately. For example, in this section, the starting state is  in  $\xi^{n+1}(j)$. Therefore, we use the functions $m_{n+1}^j(.)$ and $M_{n+1}^j(.)$ instead of $m_{n}^i(.)$ and  $M_{n}^i(.)$. Moreover, during a transition from  $\xi^{n}(i)$ to $\xi^{n+1}(j)$, the resource that has $S_{n,k(i,j)}^{i}$ is decremented by one. Thus, from $\xi^{n+1}(j)$, any resource that contains $S_{n,k(i,j)}^{i}-1$ players can be incremented by one to make the transition to $\xi^{n}(i)$ occurs.

The transition $\bZ_{n+1}(j) \rightarrow \bZ_{n}(i)$  happens with probability 
\begin{equation}
p_{\bZ_{n+1}(j) \bZ_{n}(i) }=
\begin{cases}
& \mathcal{P}_\epsilon(K) \frac{m_{n+1}^j(1)}{K} \frac{M_{n+1}^j(S_{n,k(i,j)}^{i}-1)}{N-1} \epsilon,~S_{n,k(i,j)}^{i}-1 >1\\
& \mathcal{P}_\epsilon(K)  \frac{m_{n+1}^j(1)}{K} \frac{M_{n+1}^j(1)-1}{N-1} \epsilon^2,~S_{n,k(i,j)}^{i}-1 =1.
\end{cases}
\label{eq:Znp1jToZniODL}
\end{equation}

The first line corresponds to the probability that an alone player experiments on a resource with $S_{n,k(i,j)}^{i}-1>1$ players, and that it accepts the decrease in utility. The second line corresponds to the probability that an alone player experiments on a resource with $S_{n,k(i,j)}^{i}-1=1$ player, and that both accept the decrease in utility. From the experimenter point of view, there are  $M_{n+1}^j(1)-1$ resources with one player.

Note that $p_{\bZ_{n+1}(j) \bZ_{n}(i) }$ is the term $i$ of the sum that gives $p_{\bZ_{n+1}(j)\bZ_n}=\sum_{i}p_{\bZ_{n+1}(j) \bZ_{n}(i) }$ in  \eqref{eq:probaZntoZnm1ODL}. Therefore, the first and second line of \eqref{eq:Znp1jToZniODL} are, with respect to the right indices changes, the term $i$ of the sum that gives the first term and the second term of \eqref{eq:probaZntoZnm1ODL} respectively.  These similarities are used in what follows.

The transition probabilities $p_{\bZ_{n+1}(j)\xi_3^n(i)}$, $p_{\xi_1^{n+1}(j)\bZ_{n}(i)}$, $p_{\xi_1^{n+1}(j)\xi_3^{n}(i)}$, $p_{\xi_2^{n+1}(j)\xi_1^{n}(i)}$, $p_{\xi_2^{n+1}(j)\xi_2^{n}(i)}$ and $p_{\xi_2^{n+1}(j)\xi_3^{n}(i)}$ correspond to the term $i$ of the sum that gives $p_{\bZ_{n+1}(j)\xi_3^n}$ $p_{\bZ_{n+1}(j)\bZ_{n}}$, $p_{\xi_1^{n+1}(j)\bZ_{n}}$, $p_{\xi_1^{n+1}(j)\xi_3^{n}}$,  $p_{\xi_2^{n+1}(j)\xi_1^{n}}$ and $p_{\xi_2^{n+1}(j)\xi_3^{n}}$ respectively. After changing the indices $n+1$ into $n$, $n$ into $n-1$ and $j$ into $i$, one can realize that these probabilities have already been computed. They correspond to $p_{\bZ_{n}(i)\xi_3^{n-1}}$, $p_{\bZ_{n}(i)\bZ_{n-1}}$, $p_{\xi_1^{n}(i)\bZ_{n}}$, $p_{\xi_1^{n}(i)\xi_3^{n-1}}$,  $p_{\xi_2^{n}(i)\xi_1^{n-1}}$ and $p_{\xi_2^{n}(i)\xi_3^{n-1}}$ from \eqref{eq:Zntoxi3nm1ODL}, \eqref{eq:ProbaXi1toZnm1ODL}, \eqref{eq:Probaxi1toxi3nm1m1ODL},\eqref{eq:xi2toZnm1ODL}, \eqref{eq:xi2np1toxi1nODL} and \eqref{eq:xi2ntoxi3nm1ODL} respectively. Thus, to obtain $p_{\bZ_{n+1}(j)\xi_3^n(i)}$, $p_{\xi_1^{n+1}(j)\bZ_{n}(i)}$, $p_{\xi_1^{n+1}(j)\xi_3^{n}(i)}$, $p_{\xi_2^{n+1}(j)\xi_1^{n}(i)}$, $p_{\xi_2^{n+1}(j)\xi_2^{n}(i)}$ and $p_{\xi_2^{n+1}(j)\xi_3^{n}(i)}$, we use previous probabilities by changing the indices appropriately and then, the $i$ th term of the sum that results in $N-M_{n+1}^j(0)-M_{n+1}^j(1)$ is selected.  This term corresponds to $M_{n+1}^j(S_{n,k(i,j)}^{i}-1)$ as
\begin{equation}
\sum_{i,S_{n,k(i,j)}^{i}-1>1}M_{n+1}^j(S_{n,k(i,j)}^{i}-1)=N-M_{n+1}^j(0)-M_{n+1}^j(1).
\label{eq:sumTotalclustersODL}
\end{equation}

With these modifications and, using \eqref{eq:Zntoxi3nm1ODL}, the transition $\bZ_{n+1}(j) \rightarrow \xi_3^n(i)$ has a probability

\begin{equation}
p_{\bZ_{n+1}(j) \xi_3^{n}(i)}=
\begin{cases}
& 0,~S_{n,k(i,j)}^{i}-1 >1\\
& P_\epsilon(K) \frac{m_{n+1}^j(1)}{K} \frac{M_{n+1}^j(1)-1}{N-1}2\epsilon(1-\epsilon),~S_{n,k(i,j)}^{i}-1 =1.
\end{cases}
\label{eq:Znp1jToxi3niODL}
\end{equation}

From state $\xi_1^{n+1}(j)$ it is possible to go to $\bZ_{n}(i)$ and $\xi_3^n(i)$. During transition $\xi_1^{n+1}(j) \rightarrow \bZ_{n}(i)$ the discontent player chooses a frequency that contains $S_{n,k(i,j)}^{i}-1$ players. It happens with the following probabilities
\begin{equation}
p_{\xi_1^{n+1}(j) \bZ_{n}(i)}=
\begin{cases}
&\frac{M_{n+1}^j(S_{n,k(i,j)}^{i}-1)}{N} \epsilon,~S_{n,k(i,j)}^{i}-1 >1 \\
&\frac{ M_{n+1}^j(1)-1}{N} \epsilon^2,~S_{n,k(i,j)}^{i}-1 =1.
\end{cases}
\label{eq:xi1np1jtozniODL}
\end{equation} 
The first line is similar to the term $i$ of the sum that results in the first term of \eqref{eq:ProbaXi1toZnm1ODL}. The second line corresponds to the second term of \eqref{eq:ProbaXi1toZnm1ODL}.

With the same reasoning, using \eqref{eq:Probaxi1toxi3nm1m1ODL}, $\xi_1^{n+1}(j) \rightarrow \xi_3^n(i)$ happens with probability
\begin{equation}
p_{\xi_1^{n+1}(j) \xi_3^{n}(i)}=
\begin{cases}
& 0,~S_{n,k(i,j)}^{i}-1 >1\\
&\frac{M_{n+1}^j(1)-1}{N}2 \epsilon (1-\epsilon),~S_{n,k(i,j)}^{i}-1 =1.
\end{cases}
\label{eq:xi1np1jToxi3niODL}
\end{equation}



The state $\xi_2^{n+1}(j)$ is connected to $\bZ_{n}(i)$,  $\xi_1^{n}(i)$, $\xi_2^{n}(i)$ and $\xi_2^{n}(i)$.  The probability of transition $\xi_2^{n+1}(j)\rightarrow \bZ_{n}(i)$ is obtained using \eqref{eq:xi2toZnm1ODL} as follows
\begin{equation}
p_{\xi_2^{n+1}(j) \bZ_{n}(i)}=
\begin{cases}
&   2 \frac{M_{n+1}^j(0)+2}{N}\frac{M_{n+1}^j(S_{n,k(i,j)}^i-1)}{N} \epsilon, \text{ if } C_{n,k(i,j)}^i-1>1,\\
&   \frac{M_{n+1}^j(0)+2}{N}\frac{M_{n+1}^j(1)-1}{N}\epsilon^2, \text{ if } S_{n,k(i,j)}^i-1=1.
\end{cases}
\label{eq:xi2np1zniODL}
\end{equation}
The first line is similar to the term $i$ of the sum that results in the second term of \eqref{eq:xi2toZnm1ODL}. The second line is similar to the first term of \eqref{eq:xi2toZnm1ODL}. 

The probability of transition $\xi_2^{n+1}(j)\rightarrow \xi_1^{n}(i)$, noted $p_{\xi_2^{n+1}(j)\xi_1^{n}(i)}$, is given by
\begin{equation}
\begin{cases}
& 2 \frac{M_{n+1}^j(S_{n,k(i,j)}^i-1)}{N} \epsilon\frac{N-M_{n+1}^j(1)-M_{n+1}^j(0)}{N} (1-\epsilon),  S_{n,k(i,j)}^i-1>1,\\
& 2 \frac{M_{n+1}^j(1)-2}{N} \epsilon^2 \frac{N-M_{n+1}^j(1)+1-M_{n+1}^j(0)}{N} (1-\epsilon),  S_{n,k(i,j)}^i-1=1.\\
\end{cases}
\label{eq:xi2np1jxi1iODL}
\end{equation}
which is obtained from equation \eqref{eq:xi2np1toxi1nODL}. 

The probability of transition $\xi_2^{n+1}(j)\rightarrow \xi_2^{n}(i)$ is given by
\begin{equation}
p_{\xi_2^{n+1}(j)\xi_2^{n}(i)}=
\begin{cases}
& 2\frac{M_{n+1}^j(S_{n,k(i,\ell)}^i-1)}{N}\epsilon \frac{M_{n+1}^j(1)-2}{N}(1-\epsilon)^2, \text{ if } S_{n,k(i,j)}^i-1\geq 2,\\
&\frac{M_{n+1}^j(1)-2}{N} \epsilon^2 \frac{M_{n+1}^j(1)-3}{N} (1-\epsilon)^2, \text{ if } S_{n,k(i,j)}^i-1=1.
\end{cases}
\label{eq:xi2np1jxi2niODL}
\end{equation}
which is obtained from equation \eqref{eq:xi2ntoxi2nm1ODL}.

Finally, the probability of transition $\xi_2^{n+1}(j)\rightarrow \xi_3^{n}(i)$ is obtained using equation~\eqref{eq:xi2ntoxi3nm1ODL} as follows
\begin{equation}
p_{\xi_2^{n+1}(j)\xi_3^{n}(i)}=
\begin{cases}
& 0, \text{ if } S_{n,k(i,j)}^i-1\geq 2,\\
&\frac{M_{n+1}^j(0)+2}{N}\frac{M_{n+1}^j(1)-1}{N}2 \epsilon (1-\epsilon), \text{ if } S_{n,k(i,j)}^i-1=1.
\end{cases}
\label{eq:xi2np1jxi3niODL}
\end{equation}
%
%

\ifCLASSOPTIONcaptionsoff
  \newpage
\fi



%
%
%
\bibliography{IEEEabrv,biblio_conv_journal}
\bibliographystyle{IEEEtran}
%
%
%
%
%




\end{document}